\PassOptionsToPackage{unicode}{hyperref}
\PassOptionsToPackage{naturalnames}{hyperref}
\documentclass[twocolumn,pra,letterpaper,superscriptaddress]{revtex4-1}
\usepackage{hyperref}
\usepackage{caption,float,latexsym,amsmath,amssymb,amsfonts,graphicx,color,amsthm}
\usepackage{enumerate, subfigure}
\usepackage[dvipsnames]{xcolor}
\usepackage{braket}

\allowdisplaybreaks
\begin{document}\sloppy

\title{Differential phase encoded measurement-device-independent quantum key distribution}
	\author{Shashank Kumar Ranu} %
	\affiliation{Department of Electrical Engineering, Indian Institute of Technology Madras, Chennai, India}
	\affiliation{Department of Physics, Indian Institue of Technology Madras, Chennai, India}
	\author{Anil Prabhakar}%
	\affiliation{Department of Electrical Engineering, Indian Institute of Technology Madras, Chennai, India}
	\author{Prabha Mandayam} %
	\affiliation{Department of Physics, Indian Institue of Technology Madras, Chennai, India}

\begin{abstract}
We present a measurement-device-independent quantum key distribution (MDI-QKD) using single photons in a linear superposition of three orthogonal \emph{time-bin} states, for generating the key. The orthogonal states correspond to three distinct paths in the delay line interferometers used by two (trusted) sources. The key information is decoded based on the measurement outcomes obtained by an untrusted third party Charles, who uses a beamsplitter to measure the phase difference between pulses traveling through different paths of the two delay lines. The proposed scheme combines the best of both differential-phase-shift (DPS) QKD and MDI-QKD. It is more robust against phase fluctuations, and also ensures protection against detector side-channel attacks. We prove unconditional security by demonstrating an equivalent protocol involving shared entanglement between the two trusted parties. We show that the secure key rate for our protocol compares well to existing protocols in the asymptotic regime. For the decoy-state variant of our protocol, we evaluate the secure key rate by using a phase-post-selection technique. Finally, we estimate the bit error rate and the phase error rate, in the finite key regime.
\end{abstract}
\maketitle
\section{Introduction}
\label{intro}
Quantum key distribution (QKD) is proven to be unconditionally secure in theory~\cite{mayers2001unconditional,lo1999unconditional,shor2000simple,gottesman2004security,biham1996quantum}. However, QKD protocols may be rendered insecure in practice, because of the difference in the behavior of practical devices and the respective theoretical models used in security proofs. For example, the standard protocols and their security proofs fail to take into account side-channel attacks on the detectors~\cite{Qi2007,fung2007phase, lamas2007breaking,zhao2008quantum,nauerth2009information, xu2010experimental,lydersen2010hacking,gerhardt2011full, weier2011quantum,ko2016informatic}, thereby compromising security. 
	
	Various solutions have been proposed to counteract side-channel attacks. One solution is to develop precise mathematical models of devices used in the QKD experiments and incorporate these models into new security proofs \cite{fung2008security, maroy2010security,kang2020measurement}. However, the complex nature of devices makes this approach very challenging to realize in practice. The other solution is to develop counter measures against known side-channel attacks \cite{da2012real,yuan2011resilience}, but the QKD system still remains vulnerable to unanticipated attacks. Device independent QKD (DI-QKD)\cite{mayers1998quantum,biham2006proof} is another viable candidate against side-channel attacks. The security of DI-QKD relies on the violations of Bell inequality. However, the requirement of a loophole-free Bell test, and an extremely low key rate at long distances, makes this unfeasible with current technology~\cite{gisin2010proposal,PhysRevA.84.010304}. Measurement-device-independent QKD (MDI-QKD)~\cite{lo2012measurement,braunstein2012side} was introduced as a practical solution to side-channel attacks on the measurement unit. 
	
In an MDI-QKD protocol, Alice and Bob encode their respective classical key bits into quantum states and send it to a potentially untrusted party, Charles. It is assumed that the measurement unit is under complete control of Charles, who carries out the measurement and announces the results. This is followed by sifting, error correction and privacy amplification, as carried out in standard QKD protocols. The first MDI-QKD scheme was designed for a polarization-based implementation of BB84~\cite{lo2012measurement}. Various variants of the original polarization-based MDI protocol exist in the literature \cite{wang2016enhanced,wang2018new,zhu2016enhanced}. MDI protocols employing time-bin~\cite{rubenok2013real,liu2013experimental} and phase-based encodings~\cite{tamaki2012phase,ferenczi2013security,ma2012alternative,lin2018simple,ma2018phase} also exist in the literature -- see~\cite{xu2014measurement} for a recent review. However, random phase and polarization fluctuations are a major hindrance in long distance implementations of polarization and phase-based MDI-QKD schemes. 
	
Here, we propose a differential-phase-shifted MDI-QKD (DPS MDI-QKD) scheme, as a potential candidate for alleviating random phase fluctuations. Random polarization fluctuations that occur over milli-second timescales do not affect such a differential phase-based protocol. In a differential phase encoded QKD protocol, the classical key is encoded in the phase difference between successive optical pulses which are a few nano-seconds apart, thus making the protocol resilient to the effects of environmental phase fluctuations. There are a few variants of differential-phase-shifted keying proposed in the literature~\cite{dpsreview_2015}. For example, the sender Alice could use a phase modulator in combination with a random number generator to apply a phase of either $0$ or $\pi$, randomly, on a sequence of successive pulses generated by a weak coherent source (WCS)~\cite{inoue2003differential}. Alternately, the phase modulation may be done on a single photon pulse converted into a superposition of three orthogonal states corresponding to three different time-bins, via a delay line interferometer~\cite{inoue2002differential}. 
	
	Here, we make use of the $3$-pulse protocol, whose security is based on the fact that the eavesdropper has to distinguish  between a set of four non-orthogonal quantum states. While the coherent-state DPS protocol is provably secure against individual attacks~\cite{coherent_security06}, the single-photon based $3$-pulse protocol is shown to be unconditionally secure~\cite{unconditional_dps09}. However, this security proof assumes infinitely long keys whereas experimental implementations are constrained by the finite computational power of Alice and Bob, resulting in keys of finite length.
	
	Effect of the finiteness of the key size on security parameters was first studied in~\cite{inamori2007unconditional}. Subsequently, the security of BB84~\cite{tomamichel2012tight} and decoy state protocols \cite{hayashi2014security,PhysRevA.89.022307,wang2019finite} against collective attacks in the finite-key regime was established. Techniques used for the finite-key analysis of conventional QKD have also been applied to MDI-QKD, but for specific attacks~\cite{ma2012statistical}. More recently, a rigorous security proof of MDI-QKD against general attacks for a finite key length was demonstrated~\cite{curty2014finite}.
	
	In this paper, we present a MDI-QKD scheme which incorporates the advantages of differential phase encoding. We show unconditional security of our protocol by mapping it to an equivalent entanglement-based protocol. An upper bound for the phase error rate of our scheme, in terms of the bit error rate, is then used to carry out the asymptotic and finite-key analysis of our scheme. We demonstrate that our protocol generates secure keys over reasonable distances, even under system imperfections. We also propose a decoy-state variant of our protocol and use phase-post-selection technique to show that our scheme offers reasonable security, thereby making it an attractive choice for practical implementations that use a weak coherent source (WCS).
	
	In Sec.~\ref{sec:prelim}, we briefly review the $3$-pulse DPS-QKD protocol and its security aspects. We discuss our DPS-MDI protocol in Sec.~\ref{sec:dps-mdi} and show that it maps to an entanglement-based protocol. We obtain the secure key rate using an ideal single-photon source as well as a WCS for the protocol. Finally, we present the finite-key analysis of our scheme in Sec.~\ref{sec:finite_key}. The details of the calculation of the secure key rates for our scheme, and the mapping of our protocol to an equivalent entanglement-based protocol are presented in Appendix ~\ref{sec:BMS} and~\ref{sec:EB}, respectively. We explicitly calculate the phase error rate for our protocol in terms of the bit error rate in Appendix ~\ref{sec:bound}, and finally, calculate the parameters involved in the asymptotic key analysis in Appendix ~\ref{sec:asymptotic}. 
\begin{widetext}
		\begin{center}
			\begin{figure}[h!]
				\centerline{\includegraphics[width=0.77\textwidth,height=\textheight,keepaspectratio]{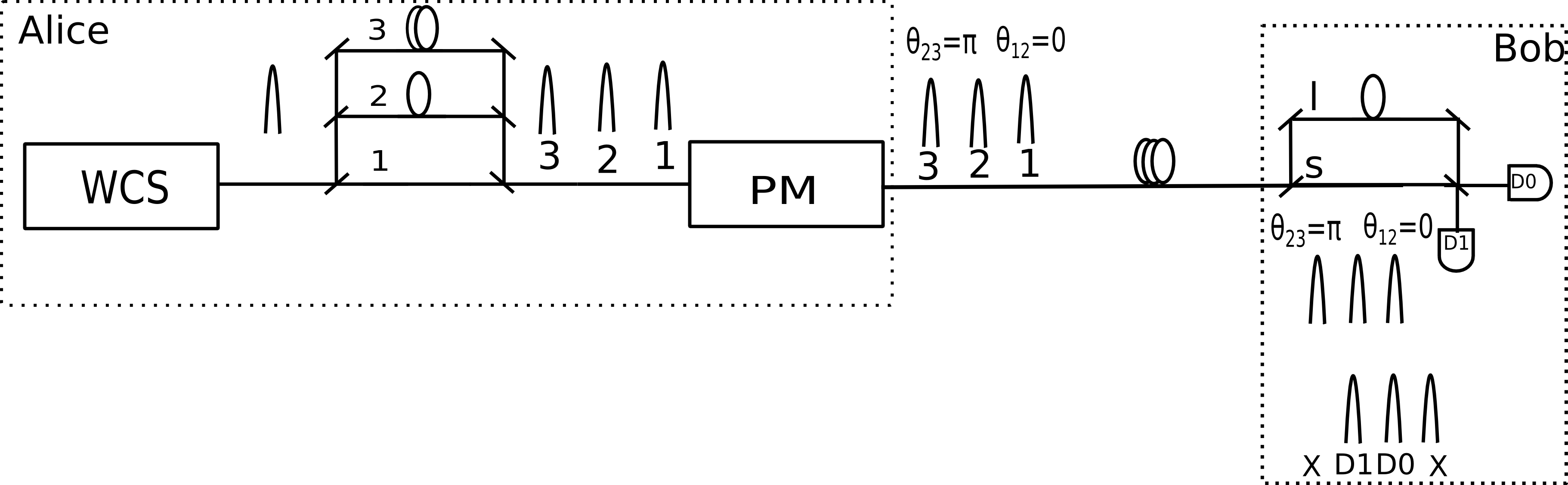}}
				\caption{Schematic of $3$-pulse differential-phase-shift QKD. WCS = Weak Coherent Source, PM = Phase Modulator, $D0, D1$ = Single-photon Detectors.} 
				\label{fig:DPS}
			\end{figure}
		\end{center}
\end{widetext}	
	\section{Preliminaries}\label{sec:prelim}
	
Starting with the original proposal to implement the  B92 protocol\cite{bennett1992quantum}, differential phase or distributed-phase protocols have been well studied in the QKD literature~\cite{dpsreview_2015}. Such protocols are popular because they are relatively easy to implement (compared to polarization-based protocols) and are robust against phase fluctuations. Most phase-based schemes use weak coherent pulses for encoding the key. However, in this paper, we use the single-photon scheme proposed in~\cite{inoue2002differential}. We shall henceforth refer to this scheme as the $3$-pulse DPS-QKD protocol and provide a brief description below.
	
	\subsection{$3$-pulse differential-phase-shift keying}~\label{sec:3pulse}
	
In a $3$-pulse DPS-QKD protocol, the sender (Alice) throws a single photon into a superposition of three time-bins, corresponding to the three distinct paths of a delay line interferometer, and then uses a phase modulator to introduce a relative phase between successive time-bins, as shown schematically in Fig.~\ref{fig:DPS}. Alice encodes her random key bit $\{0,1\}$ as a random phase $\{0,\pi\}$ between successive pulses. The receiver (Bob) thus gets one of the four non-orthogonal quantum states given below, corresponding to the four possible phase-differences, i.e.,
	\begin{eqnarray}
	 |\psi (\pm, \pm) \rangle = \frac{1}{\sqrt{3}}\left(\, |100\rangle _{a} \pm |010\rangle_{a}  \pm |001\rangle_{a} \,\right) . \label{eq:3pulse_dps}
	\end{eqnarray}
	
	Here, $\ket{100}_a$, $\ket{010}_a$ or $\ket{001}_a$ indicate that the photon travelled with equal probability via paths 1, 2 or 3, respectively, in Alice's set-up.
		
	Bob's decoding setup comprises of a delay line interferometer (DLI) and two single-photon detectors. The path lengths are chosen such that the longer arm of Bob's DLI introduces a time delay $\Delta t$ which is exactly equal to the difference in time taken by the photon to traverse  two successive arms of Alice's  3-path delay line. Thus, Bob can detect the incoming photon in one of the four possible time-bins, which we label as $t_{1}$, $t_{2}$, $t_{3}$, $t_{4}$, each separated from its previous bin by a time of $\Delta t$. Detections at times $t_{1}$ and $t_{4}$ do not provide any phase information, whereas detections at times $t_{2}$ and $t_{3}$ provide information about the relative phases $\theta_{12}$ and $\theta_{23}$ respectively (see Fig.~\ref{fig:DPS}). Specifically, Bob decodes the key bit associated with a given time-slot as a $0$, or $1$, if detector $D0$, or $D1$, clicks. By publicly announcing his detection times, Bob performs key-reconciliation with Alice and it is easy to see that the sifted key rate for  this $3$-pulse protocol is $1/2$. 
	
	An alternate form of phase-encoded QKD is the pulse-train DPS-QKD~\cite{inoue2003differential}, which is a variant of the original B92 protocol~\cite{bennett1992quantum}. In the pulse-train protocol, Alice generates a train of coherent pulses and applies a phase of $0$ or $\pi$ to the pulses randomly, to encode the key bits $0$ or $1$, respectively. These phase modulated pulse trains are sent to Bob, who passes the incoming pulses through a DLI. Depending upon the phase difference between two successive pulses, constructive or destructive interference happens. An MDI-QKD protocol based on the coherent-state pulse-train DPS protocol was also proposed in ~\cite{ferenczi2013security}.

	We refer to~\cite{shashank2017DPS} for a detailed analysis of the secure key rate for the $3$-pulse DPS protocol, assuming individual attacks. A simple comparison with the pulse-train DPS protocol~\cite{diamanti2006security} shows that the $3$-pulse variant offers better security against individual attacks, in the following sense: an eavesdropper introduces a higher error rate and also has a lower learning rate in the $3$-pulse protocol~\cite{shashank2017DPS}.\\
		Finally, we note that the $3$-pulse DPS-QKD protocol can be extended to an $n$-pulse protocol by increasing the number of possible paths that the single photon can take at the sender's set-up. In fact, the single photon DPS protocol using $n$ such paths has been shown to be unconditionally secure against general attacks for any $n\geq 3$~\cite{unconditional_dps09}. Experimental realization of $n$ path DPS protocol would involve generating a photon in a superposition of $n$ paths/time-bins using passive beam splitters (or beam combiners). As we increase the number of paths/time-bins, the insertion loss of passive beam splitters reduces the sifted key rate by a factor of $n$. Scaling of $n$ in an experimental realization thus reduces the sifted key rate, in fact, the $n=3$ protocol is shown to achieve the optimal secure key rate per pulse~\cite{unconditional_dps09,shaw2020equivalence}. Note that, $n=3$ is the smallest $n$ that allows Alice and Bob to encode the key information in a non-orthogonal set of states using only two phase values, $0$ and $\pi$. 
	
	
	
	%
	%
	\section{DPS-MDI-QKD}\label{sec:dps-mdi}
	We now describe our MDI-QKD protocol based on the $3$-pulse phase encoding scheme, using an ideal single-photon source. Apart from the fact that this scheme offers better security against individual attacks, compared to other DPS protocols, there are other practical considerations that motivate our use of the $3$-path superposition in our protocol. 
	\begin{enumerate}
		\item  When Alice and Bob both use an ideal-single-photon source to implement a pulse-train protocol using two phase values ($0$ and $\pi$) for encoding key bits, the phase-independent nature of Hong-Ou-Mandel interference~\cite{ou2007multi} makes the key extraction difficult.  
		\item Using only two phase values ($0$ and $\pi$) makes the states in a two-pulse protocol orthogonal, making them perfectly distinguishable~\cite{unconditional_dps09}.
	\end{enumerate}
	
	Hence, we need at least $3$-paths in the superposition to implement an MDI protocol using only a pair of phases ($0$, $\pi$) for the encoding. An MDI protocol based on a two-path superposition, and four phase values $(0,\frac{\pi}{2},\pi,\frac{3\pi}{2})$ was proposed in~\cite{ma2012alternative}. This scheme yields a phase-encoded version of BB84, with a sifted key rate of 1/2, but it needs four different voltage levels for driving the phase modulator in order to encode the key information. Now, increasing the number of voltage levels in a high speed phase modulator driver circuit leads to an increase in amplitude fluctuations, consequently increasing the quantum bit error rate \cite{korzh2013high}. Our proposed DPS MDI protocol reduces the complexity of the key encoding process by using only two phase values ($0$, $\pi$), with a sifted key rate of 4/9, explained in section ~\ref{subsection:reconcile} below.  
	
	A simple schematic is shown in Fig.~\ref{fig1}. As before, Alice and Bob generate single-photon pulses that pass through their respective DLIs, each creating the superposition state described in Eq.~\eqref{eq:3pulse_dps}. Alice and Bob then encode the random key bits $\{0,1\}$ by assigning a relative phase difference of $\{0,\pi\}$ between two successive pulses, and send their encoded signal states to the measurement unit (Charles). Charles' measurement set-up comprises of a beamsplitter and two single-photon detectors, labeled $D_{c}$ and $D_{d}$ as indicated in Fig.~\ref{fig1}. For every photon detected by his setup, he notes which detector clicked ($D_{c}$ or $D_{d}$ or both), and the corresponding time-bin ($t_{1}$, $t_{2}$ or $t_{3}$) at which the click was observed. Based on this information, which is made public by Charles, Alice and Bob extract a sifted key. 
\subsection{Sifting and Reconciliation}\label{subsection:reconcile}
We may use the form of the encoded $3$-pulse state in Eq.~\eqref{eq:3pulse_dps} to represent the input to the Charles' measurement module as,
\begin{widetext}
		\begin{center}
			\begin{figure}[h!]
				\centerline{\includegraphics[width=1\textwidth,height=\textheight,keepaspectratio]{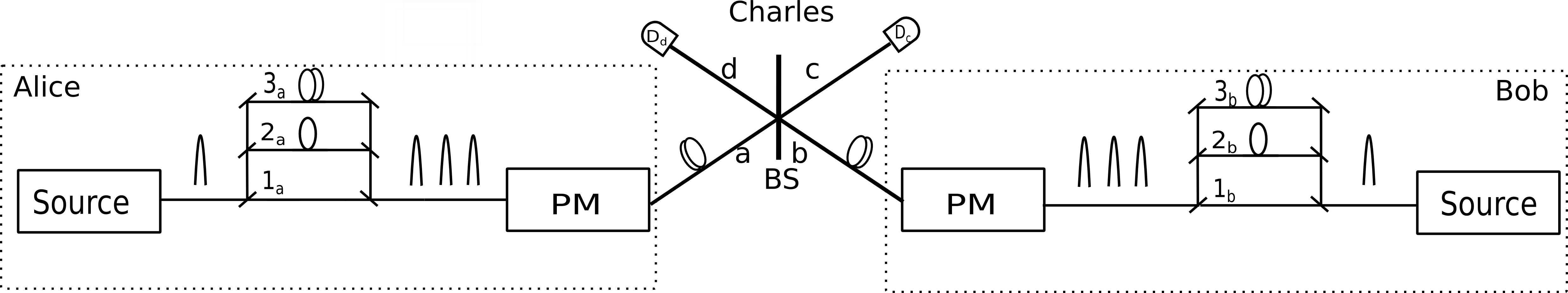}}
				\caption{Schematic of differential phase encoded MDI-QKD. Here, PM = Phase Modulator, $D_{c},D_{d}$ = Single -photon detectors.}
				\label{fig1}
			\end{figure}
		\end{center}
		%
		%
		\begin{eqnarray} 
		\ket{\psi (\phi_{a_{1}}, \phi_{a_{2}}, \phi_{b_{1}}, \phi_{b_{2}} ) }_{\text{in}}&=& \frac{1}{\sqrt{3}} \left( \, \ket{100}_{a} + e^{i\phi_{{a}_{1}}} \ket{010}_{a} + e^{i\phi_{{a}_{2}}} \ket{001}_{a} \, \right) 
		\otimes \frac{1}{\sqrt{3}} \left( \, \ket{100}_{b} + e^{i\phi_{{b}_{1}}} \ket{010}_{b} + e^{i\phi_{{b}_{2}}} \ket{001}_{b} \, \right) . \label{eq:dps_mdi}
		\end{eqnarray}
 \end{widetext}

	As before, $1$ and $0$ indicate the presence or absence of a photon in a particular path. Similarly, $\ket{100}_{a}$ is the $3$-pulse superposition state corresponding to the photon traversing path $1_{a}$ in Alice's set-up, $\ket{010}_{b}$ is a $3$-pulse state corresponding to photon traversing path $2_{b}$ in Bob's set-up, and likewise for other terms in Eq.~\eqref{eq:dps_mdi}. For the sake of brevity, we represent tensor products of the form $\ket{100}_{a} \otimes \ket{100}_{b}$ as $\ket{100,100}_{ab}$ in the rest of the paper. In DPS-MDI, Alice and Bob encode classical information as phase differences between first and second time-bins, and second and third time-bins. In our analysis, we assume the phase of the first time-bin as the reference phase, and apply a suitable phase ($0$ or $\pi$ relative to the reference phase) on the second and third time-bins to encode the key information.
	
Corresponding to every pair of photons generated by the sources, there are three distinct time-bins ($t_{1}, t_{2}, t_{3}$) at which Charles' detectors click, corresponding to paths $1_{a}, 2_{a}, 3_{a}$ and $1_{b}, 2_{b}, 3_{b}$ in Alice's and Bob's set-up respectively. We first rewrite Charles' input state by grouping pairs of pulses that arrive in the same time-bin:
\begin{widetext}
				\begin{eqnarray} 
						&& \ket{\psi}_{\text{in}} = \frac{1}{3}[ \; \ket{100,100}_{ab} + e^{i\phi_{{a}_{1}}}\ket{010,100}_{ab}+ e^{i\phi_{{a}_{2}}}\ket{001,100}_{ab}+ e^{i\phi_{{b}_{1}}} \ket{100,010}_{ab} + e^{i\phi_{{b}_{2}}}\ket{100,001}_{ab}  \nonumber \\
						&&  + e^{i(\phi_{{a}_{1}}+\phi_{{b}_{1}})}\ket{010,010}_{ab} + e^{i(\phi_{{a}_{1}}+\phi_{{b}_{2}})} \ket{010,001}_{ab}+ e^{i(\phi_{{a}_{2}}+\phi_{{b}_{1}})} \ket{001,010}_{ab}+ e^{i(\phi_{{a}_{2}}+\phi_{{b}_{2}})} \ket{001,001}_{ab} \; ].\label{eq:dps_mdi2}
						\end{eqnarray}
	\end{widetext}
	Note that the pairs of photons that traverse through identical paths in Alice's and Bob's interferometer (such as $(1_{a}, 1_{b})$ or $(2_{a}, 2_{b})$ or $(3_{a}, 3_{b})$) do not contribute to the sifted key. Such a pair of photons would bunch together due to Hong-Ou-Mandel interference~\cite{ou2007multi} and come out at the same port of the beamsplitter. 
	
	Using a beamsplitter transformation, we can write down the final two-photon state after the action of Charles' beamsplitter. We refer to Appendix~\ref{sec:BMS} for the details of the calculation, with the form of the final state after Charles' measurement given in Eq.~\eqref{eq:bsm_output}. We observe that depending on the values of the relative phases $\Delta \phi_{1} = \phi_{a_{1}}-\phi_{b_{1}}$  and $\Delta \phi_{2} = \phi_{a_{2}}-\phi_{b_{2}}$, and the path traversed by Alice's and Bob's photons, Charles may have the same or different detectors click at two different time-bins. 
	
	Finally, Alice and Bob perform key reconciliation once Charles announces his measurement outcomes.  Based on which detector ($D_{c}$ or $D_{d}$) clicks and the time-bins ($t_{1}$, $t_{2}$ and $t_{3}$) corresponding to the clicks for each pair of signal states, Alice and Bob can generate the sifted key using either $\Delta\phi_{1}$ or $\Delta \phi_{2}$ as listed in Table I. 
	
	\begin{table*}  
		Table I : Key reconciliation scheme for the proposed protocol. We write $\frac{1}{\sqrt{2}}[\ket{0}_{A_{1}}\ket{0}_{B_{1}} -\ket{1}_{A_{1}}\ket{1}_{B_{1}}]$ as $\frac{1}{\sqrt{2}}[\ket{00}_{A_{1}B_{1}} -\ket{11}_{A_{1}B_{1}}]$, and similarly for other terms. 
		\centering 
		\begin{tabular}{|c|c|c|c|} 
			\hline 
			\text Measurement outcome of Charles & \text Action of Alice and Bob & \text Requirement of bit flip & \text Shared EPR pair \\ [0.2 cm] \hline\hline
			Det c clicks at both $t_{1}$ and $t_{2}$ & Extract key using $\Delta\phi_{1}$ & No &$\frac{1}{\sqrt{2}}[\ket{00}_{A_{1}B_{1}} -\ket{11}_{A_{1}B_{1}}]$ \\[0.2 cm]\hline
			Det d clicks at both $t_{1}$ and $t_{2}$ & Extract key using $\Delta\phi_{1}$ & No &$\frac{1}{\sqrt{2}}[\ket{00}_{A_{1}B_{1}} -\ket{11}_{A_{1}B_{1}}]$\\[0.2 cm]\hline
			Det c clicks at both $t_{1}$ and $t_{3}$ & Extract key using $\Delta\phi_{2}$ & No &$\frac{1}{\sqrt{2}}[\ket{00}_{A_{2}B_{2}} -\ket{11}_{A_{2}B_{2}}]$\\[0.2 cm]\hline
			Det d clicks at both $t_{1}$ and $t_{3}$ & Extract key using $\Delta\phi_{2}$ & No &$\frac{1}{\sqrt{2}}[\ket{00}_{A_{2}B_{2}} -\ket{11}_{A_{2}B_{2}}]$\\[0.2 cm]\hline
			Det c clicks at $t_{1}$ and det d at $t_{2}$ & Extract key using $\Delta\phi_{1}$ & Yes &$\frac{1}{\sqrt{2}}[\ket{01}_{A_{1}B_{1}} -\ket{10}_{A_{1}B_{1}}]$\\[0.2 cm]\hline
			Det c clicks at $t_{2}$ and det d at $t_{1}$ & Extract key using $\Delta\phi_{1}$ & Yes &$\frac{1}{\sqrt{2}}[\ket{01}_{A_{1}B_{1}} -\ket{10}_{A_{1}B_{1}}]$ \\[0.2 cm]\hline
			Det c clicks at $t_{1}$ and det d at $t_{3}$ & Extract key using $\Delta\phi_{2}$ & Yes &$\frac{1}{\sqrt{2}}[\ket{01}_{A_{2}B_{2}} -\ket{10}_{A_{2}B_{2}}]$ \\[0.2 cm]\hline
			Det c clicks at $t_{3}$ and det d at $t_{1}$ & Extract key using $\Delta\phi_{2}$ & Yes &$\frac{1}{\sqrt{2}}[\ket{01}_{A_{2}B_{2}} -\ket{10}_{A_{2}B_{2}}]$ \\[0.2 cm]\hline
			Det c clicks at both $t_{2}$ and $t_{3}$ & Discard the bits  & - & - \\[0.2 cm]\hline
			Det d clicks at both $t_{2}$ and $t_{3}$ & Discard the bits & -  & - \\[0.2 cm]\hline
			Det c clicks at $t_{2}$ and det d at $t_{3}$ & Discard the bits & - & - \\[0.2 cm]\hline
			Det c clicks at $t_{3}$ and det d at $t_{2}$ & Discard the bits & -& -  \\[0.2 cm]\hline
			
		\end{tabular}\label{tab:sifting}
	\end{table*}

	It follows immediately that the the sifted key rate of our protocol is,
	\begin{equation}
R_{\rm sift} = \frac{2}{3}\times\frac{2}{3}=\frac{4}{9}. \label{eq:R_sift}
\end{equation}
We discard the clicks that occur when photons from Alice and Bob fall on the beam splitter in the same time-bin. The terms $\ket{100,100}_{ab}$, $\ket{010,010}_{ab}$ and $\ket{001,001}_{ab}$ in Eq.~\eqref{eq:dps_mdi2} correspond to such a scenario. Photons arriving at the same time-bin causes Hong-Ou-Mandel interference which leads to two photons falling on the same detector in the same time-bin, thereby making key extraction difficult. Comparing Eqs.~\eqref{eq:dps_mdi2} and~\eqref{eq:dps_mdi3}, we see that one-third of the incoming photons have to be discarded due to Hong-Ou-Mandel interference. This leads to the first factor of $\frac{2}{3}$. Next, we observe from the key reconciliation table that two-thirds of Charles' measurements contribute to the raw key, thus leading to a sifted key rate of $\frac{4}{9}$. We note that the MDI protocol based on $3$-pulse encoding offers a lower key rate compared to the one based on coherent-state pulse-train encoding~\cite{ferenczi2013security}. However, the use of single-photon sources in our protocol allows us to carry out a finite key analysis using the framework presented in ~\cite{scarani2008quantum}. Our protocol is also immune against eavesdropping attacks which target multi-photon pulses.
\subsection{An equivalent entanglement-based protocol}\label{sec:entanglement}
To analyse the security of DPS-MDI, 
	we first map it to a protocol that involves shared entangled pairs between Alice and Bob. Such a mapping of a phase-encoded protocol to an entanglement-based protocol has been shown earlier~\cite{unconditional_dps09}. 
	Following a similar approach, we now show there exists an equivalent, entanglement-based protocol to our proposed DPS-MDI-QKD protocol. The equivalent description of DPS-MDI, in terms of entangled states, allows us to demonstrate the unconditional security of our protocol and also perform the key rate analyses.
	
	We first represent Alice's single-photon pulse in a linear superposition of three orthogonal states, 
	\begin{equation}
	\ket{\psi}_{\text{a}}=\frac{1}{\sqrt{3}}\sum_{k=1}^{3} a_{k}^{\dagger}\ket{0} .
	\end{equation}
	Here, $a_{k}^{\dagger}$ denotes the creation operator for the photon in the $k^{\rm th}$ time-bin. Alice uses a quantum random number generator to generate a random $2$-bit integer $j$, written in binary notation as $(j_{1}j_{2})_{2}$. She encodes this random integer in the single-photon pulse, such that the encoded state is written as,
	\begin{eqnarray}
	\ket{\psi_{j_1j_2}}_{\text{a}}&=&\frac{1}{\sqrt{3}}\left( a^{\dagger}_{1}|0\rangle +(-1)^{j_{1}}a^{\dagger}_{2}|0\rangle+(-1)^{j_{2}}a^{\dagger}_{3}|0\rangle\right).
	 \label{eq:enc}
	\end{eqnarray}
	 Alice prepares and stores $2$ qubits corresponding to each encoded block in her quantum memory. She prepares $\ket{j_1}$ in $\ket{0}$ ($\ket{1}$) state when she applies a phase of $0$ ($\pi$) to her second time-bin. Similarly, she prepares $\ket{j_2}$ in $\ket{0}$ ($\ket{1}$) state when she applies a phase of $0$ ($\pi$) to her third time-bin. In this way, she entangles her two qubits to the encoded single-photon state as
	\begin{equation}
	\ket{\psi}_{\text{Alice}}=\frac{1}{2}\sum_{j_1,j_2\in\{0,1\}}\ket{j_{1}j_{2}}_{A_1A_2}\otimes\ket{\psi_{j_1j_2}}_{\text{a}}. \label{eq:alice}
	\end{equation}
	Bob also carries out a similar encoding procedure to get his own register of qubits entangled with his encoding blocks. Along the lines of Eqs.~\eqref{eq:enc} and~\eqref{eq:alice}, Bob's state is written as,
	\begin{equation}\label{eq:Bob_encoded_state}
	\ket{\psi}_{\text{Bob}}=\frac{1}{2}\sum_{\tilde{j_1},\tilde{j_2}\in\{0,1\}}\ket{\tilde{j}_{1}\tilde{j}_{2}}_{B_1B_2}\otimes\ket{\psi_{\tilde{j}_1\tilde{j}_2}}_{\text{b}},
	\end{equation}
where $\tilde{j_1}$ or $\tilde{j_1}$ are the random integers used by Bob to encode his single-photon pulse.
	
Alice and Bob send their encoded states across to Charles. He first applies a quantum non-demolition (QND) measurement to find the number of photons in a given state and throws away the ones which have more than one photon in the same time-bin. He sends the  rest through his beamsplitter.  He then publicly announces the time-bin (say $k=1,2,\,\text{or}\,3$), as well as the detector ($D_{c}$ or $D_{d}$), at which the photon was detected. As explained in Table I, based upon Charles' measurement outcome, Alice and Bob use either $\Delta\phi_{1}$ or $\Delta\phi_{2}$ to extract the key.

When their shared key is established using $\Delta\phi_{i}$, Alice and Bob retain their corresponding ancilla qubits ($A_{i}$ and $B_{i}$, respectively) and discard the other ancilla qubit. As shown in Appendix~\ref{sec:EB}, for those time slots when they do not need to carry-out a bit flip operation, they share a perfectly correlated entangled state $\frac{1}{\sqrt{2}}[\ket{00}_{A_{i}B_{i}} -\ket{11}_{A_{i}B_{i}}]$. On the other hand, corresponding to those time slots when they execute a bit-flip to extract the shared key, they share the anti-correlated Bell state $\frac{1}{\sqrt{2}}[\ket{01}_{A_{i}B_{i}} -\ket{10}_{A_{i}B_{i}}]$. Thus, Charles measurement and filtering effectively implements a Bell state measurement, thereby entangling Alice's and Bob's ancilla qubits. A detailed discussion of the joint state after Charles' measurement and key-reconciliation can be found in Appendix~\ref{sec:EB}.

	\subsection{Asymptotic Secure Key Rate}
	Alice and Bob perform classical post-processing on the sifted key to extract the final secure key from it.  The first step of this post-processing is to estimate the error rate in the sifted key, which involves Alice and Bob exposing a fraction of their sifted key bits to calculate the error rate. They abort the protocol and start again from the beginning (i.e signal transmission to Charles) if their calculated error rate exceeds a pre-defined threshold. They define this threshold error rate by taking into account the error introduced in the key, both due to the system imperfections as well as any potential eavesdropping.

When the estimated error rate lies below the threshold error rate, they carry out the second step of post-processing, i.e., error correction. Alice and Bob apply a suitable error correction scheme on their sifted key to correct all the erroneous bits. The error estimation and correction happens over a classical channel, and we must assume that Eve is privy to all the information exchanged between Alice and Bob. Therefore, the final step of post-processing is privacy amplification, which aims to reduce Eve's knowledge about the key well below an acceptable level. This is done by discarding a fraction of the error-free key. Alice and Bob typically use a hash function to carry out privacy amplification. 

Using the sifted key rate obtained in Eq.~\eqref{eq:R_sift} and following the analysis in~\cite{shor2000simple,lo2012measurement}, we obtain the following asymptotic secure key rate for our MDI-DPS protocol,
	\begin{equation}\label{eq:asym}
	 R  \geq Y_{11}[1-fh(e_{\text{b}})-h(e_{\text{p}})].
	 \end{equation}
Here, $Y_{11}$ is the probability of a successful Bell state measurement (BSM) when Alice and Bob transmit single photons. As per our mapping of DPS-MDI to an equivalent entangled-based protocol, a successful BSM corresponds to the cases tabulated in Table I where Charles' measurement outcomes contribute to the sifted key. $e_{\text{b}}$ is the quantum bit-error rate (QBER), $e_{\text{p}}$ is the phase error rate, $f$ represents the inefficiency of the error correction scheme employed by Alice and Bob, and $h(x)$ is the binary entropy function. 

We bound the phase error rate of our protocol in terms of the bit error rate in Appendix ~\ref{sec:bound} as,
\begin{equation}
e_{\text{p}}\leq e_{\text{b}}, \label{eq:main_bound}
\end{equation}
 and use this bound for all of the simulation results. We also explicitly calculate the parameters given in Eq.~\eqref{eq:asym} for our protocol in Appendix~\ref{sec:asymptotic}. We have taken phase misalignment, dark counts and different channel losses for the two channels into consideration while obtaining these parameters.
\begin{center}
\begin{figure}[htbp!]
    \centerline{\includegraphics[width=0.5\textwidth,height=\textheight,keepaspectratio]{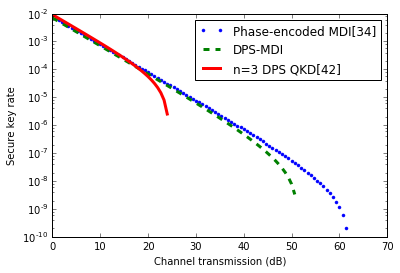}}
	 \caption{Key rates for ideal, single-photon based protocols.}
	 \label{fig:single}
\end{figure}
\end{center}
We compare the asymptotic key rate of DPS-MDI with two other protocols - phase-encoded MDI protocol ~\cite{ma2012alternative} and DPS QKD \cite{unconditional_dps09}. From Fig.~\ref{fig:single}, we observe that DPS-MDI offers a secure channel length which is nearly twice of the channel length of DPS QKD - a trademark of MDI protocols when compared with non-MDI protocols. We also see that our DPS-MDI protocol offers performance comparable to an existing phase-encoded MDI protocol in terms of secure channel length and key rate. The slightly higher key rate in \cite{ma2012alternative} is attributed to its higher sifted key rate of $\frac{1}{2}$ compared to DPS-MDI's rate of $\frac{4}{9}$.

We have obtained the non-MDI DPS QKD plot in Fig.~\ref{fig:single} by using the key rate equation derived in \cite{unconditional_dps09}. We would like to point out the difference in the secure channel length for n=3 DPS QKD obtained in \cite{unconditional_dps09} and our simulation. The difference arises because we have used $3\times 10^{-6}$ as the dark count probability in our simulation, which is $1000$ times higher than the dark count probability used in \cite{unconditional_dps09}. Also, \cite{unconditional_dps09} assumes an ideal error correction step in their classical post-processing, while our simulations assume a non-ideal error correction step. We capture the inefficiency of error correction in our protocol using the parameter $f$ (Eq. ~\ref{eq:asym}).
	 
In experimental implementations, weak coherent sources (WCS) are typically used to generate pulses with mean photon number ($\mu$) of less than one so that the probability of generation of multi-photon pulses is significantly less than that of single-photon pulses. However, a WCS could still generate multi-photon pulses, and leak information to Eve. Hence, we use the decoy-state method to establish the security of our DPS-MDI protocol. The original decoy-state based QKD protocols have been proposed for BB84 schemes and secure key rates obtained in ~\cite{Lo,Ma}. Decoy state analysis for MDI-BB84 was done in ~\cite{lo2012measurement}. In our case, we follow the approach in ~\cite{lo2012measurement} along with the improved phase-post-selection technique employed in ~\cite{ma2012alternative} to obtain the key rate as, 
	\begin{equation}\label{eq:decoy}
    R \geq Q_{11}[1-h(e_{\text{p}})]+Q^{'}_{0\mu_{b}}-I_{\text{ec}}.
	\end{equation}
    Here, $I_{\text{ec}}$ is the cost of error correction written as
  \begin{equation}\label{eq:cost}
  I_{\text{ec}}=Q_{\mu_{a}\mu_{b}}fh(E_{\mu_{a}\mu_{b}}),
  \end{equation}
  where $Q_{\mu_{a}\mu_{b}}(E_{\mu_{a}\mu_{b}})$ is the overall gain (QBER) when Alice and Bob use a WCS with mean photon numbers $\mu_{a}$ and $\mu_{b}$, respectively. $Q_{11}(e_{\text{p}})$ is the gain (phase error rate) when both the sources generate single-photon states, and $Q^{'}_{0\mu_{b}}=e^{- \mu_a}Q_{0\mu_{b}}$ is the probability that there is no photon from Alice's side and a successful BSM occurs. We refer to Appendix ~\ref{sec:asymptotic} for formal definitions and a detailed evaluation of these parameters.
  
\begin{figure}[htbp!]
 \centerline{\includegraphics[width=0.5\textwidth,height=\textheight,keepaspectratio]{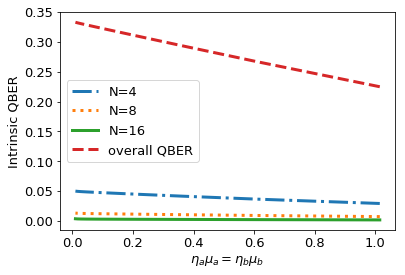}}
 \caption{QBER comparison when phase randomization is (a) carried out over entire range (b) carried out in one of the N slices. We have used Eq.~\eqref{eq:integral_qber} for numerically evaluating the QBER}
\label{fig:qber}
\end{figure}

Our decoy-state analysis assumes a fully phase-randomized coherent source. The intrinsic QBER shoots up due to phase randomization of the coherent source. The overall phase of $[0,2\pi)$ can be sliced into N distinct slices as,
\begin{eqnarray}\label{eq:slice}
    \bigg[\frac{m\pi}{N},\frac{(m+1)\pi}{N}\bigg)\cup\bigg[\frac{(m+N)\pi}{N},\frac{(m+N+1)\pi}{N}\bigg),
	\end{eqnarray}	
	where $m$ ranges from 0 to $N-1$. Instead of carrying out phase randomization over the entire interval $[0,2\pi)$, Alice and Bob randomly select one slice out of $N$, and then randomize the phase. Hence, an additional step of revealing the selected slice gets added in the decoy state version of our protocol. Alice and Bob keep the bits when both of them have selected the same phase slice. Fig. ~\ref{fig:qber} shows that dividing the interval $[0,2\pi)$ into  slices reduces the intrinsic QBER from $34\%$ to around $1\%$ for $N=16$. 
	 				
\begin{center}
  		 				\begin{figure}[t!]
  		 					\centerline{\includegraphics[width=0.5\textwidth,height=\textheight,keepaspectratio]{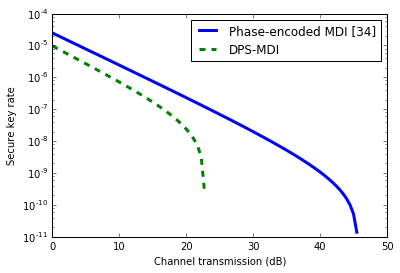}}
  		 					\caption{Key rate comparison for decoy-state MDI schemes.}
  		 					\label{fig:decoy}
  		 				\end{figure}
  	\end{center}

However, this phase-post-selection technique also changes the cost of error correction mentioned in Eq. ~\eqref{eq:cost} to
  \begin{equation}\label{eq:cost_increase}
  I_{\text{ec}}=\sum\limits_{m}Q^{m}_{\mu_{a}\mu_{b}}fh(E^{m}_{\mu_{a}\mu_{b}}).
  \end{equation}
 From our numerical simulations, we observe that the key rate becomes negative upon using Eq.~\eqref{eq:cost_increase} in conjugation with Eq.~\eqref{eq:decoy}. Hence, we assume that the gain and error rate of the single-photon states are evenly distributed over all the slices, thereby modifying the decoy MDI key rate equation to
 	\begin{equation}\label{eq:decoy_modified}
     R \geq \frac{1}{N}Q_{11}[1-h(e_{\text{p}})]+Q^{'}_{0\mu_{b}}-Q^{m}fh(E^{m})\vert_{m=0}.
 	\end{equation}
  We refer to Appendix ~\ref{sec:asymptotic} for a detailed analysis of the effect of this phase-post-selection technique on the overall gain and QBER. We compare the key rate of our decoy-state DPS-MDI with ~\cite{ma2012alternative} (see Fig.~\ref{fig:decoy}), where
 we used the parameters from ~\cite{lo2012measurement} for our simulations. The quantum efficiency of the detectors was taken to be 14.5\% with a misalignment error of 1.5\%. $N$ and $f$ are taken to be 16 and 1.16 respectively. We assume a dark count rate of $3\times10^{-6}$ for the detector and an attenuation of 0.2 dB/km in the fiber channel. 
	\subsection{Practical implementation}
	As described above, Alice and Bob can share a secure key using the setup shown in Fig.~\ref{fig1}. However, a practical implementation of the proposed scheme requires certain modifications to the set-up (see Fig.~\ref{practical}). 
	 \begin{enumerate}
	 	\item Key generation requires detection of two time-synchronized photons by a single detector. In practice, this would be constrained by the finite dead-time of a single-photon detector. Hence, an acousto-optic deflector (AOD) is used to route the photon in each time-bin to different single-photon detectors. This results in a slight modification to the key-reconciliation step, namely, Charles now announces which pair of detectors clicked in each time-bin.
			\item Alice and Bob need a common phase reference, since they use independent laser sources for generating their single-photon pulses. The optical phase-locked loop (OPLL) technique~\cite{kahn1989optical,kazovsky1985decision}, commonly used in coherent detections, can be used to phase lock the sources used by Alice and Bob. The OPLL has a simple setup and requires only off-the-shelf components~\cite{ferrero2008optical}.
\end{enumerate}

\section{Finite key analysis of DPS-MDI-QKD}\label{sec:finite_key}
Finiteness of the key size constitutes a major chink in the security proofs of practical QKD protocols. Most of the theoretical proofs provide a bound on the secure key rate by assuming the key size as infinite. However, practical implementations cannot run forever. This gap in theory and practice is bridged by providing security bounds for a finite number of signal exchanges between Alice and Bob.

A perfect key is a uniformly distributed bit string, having no dependence on an adversary's knowledge. Practical keys deviate from this ideal scenario, and this deviation is captured by a parameter $\varepsilon$, interpreted as the maximum probability of a practical key differing from a completely random bit string. Following~\cite{renner2005quantum,ben2005universal}, we say that a key $K$ is $\varepsilon$-secure with respect to an eavesdropper $E$ if,  
	\begin{equation}
		\frac{1}{2}\parallel \rho_{KE}-\tau_{K}\otimes\rho_{E}\parallel_{1}\;\leq\; \varepsilon. 
\end{equation}
				\begin{widetext}
			\begin{center}
				\begin{figure}[h!]
					\centerline{\includegraphics[width=\textwidth,height=\textheight,keepaspectratio]{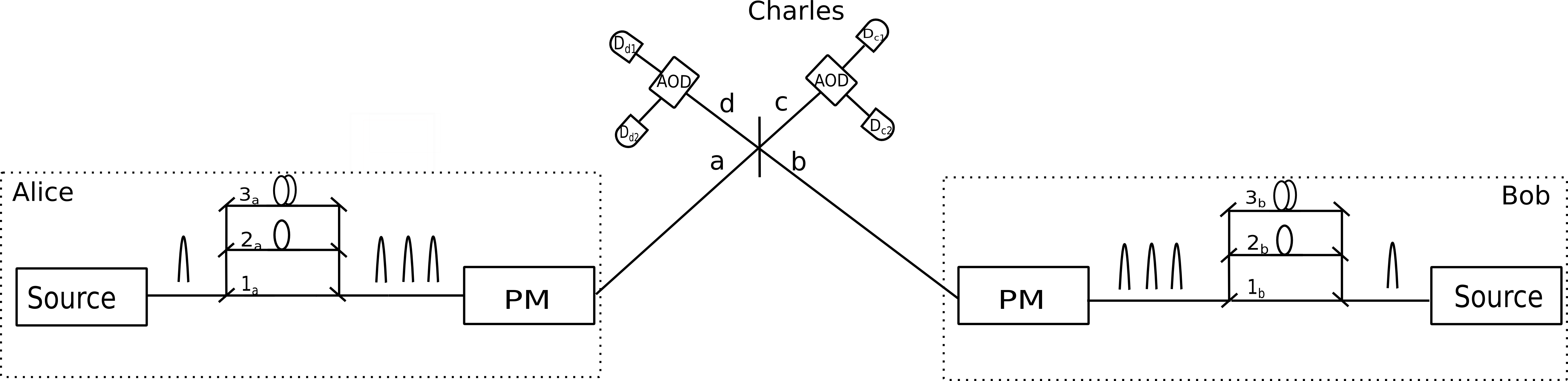}}
					\caption{Schematic of a practical $3$-pulse DPS-MDI-QKD implementation. Alice and Bob use a phase modulators (PM) and a delay line interferometer each. Charlie's set-up comprises a beamsplitter, four detectors and two acousto-optic deflectors (AODs).}
					\label{practical}
				\end{figure}
			\end{center}
			\end{widetext}

Here, $\rho_{KE}$ is the joint state of the `key system' $K$ and the adversary $E$, $\rho_{E}$ is the state held by the adversary, and $\tau_{K}$ is the completely mixed state on $K$.
	
In the asymptotic case, for any QKD protocol where Alice and Bob share entangled pairs, the secure key rate ($R$) can be bounded under the assumption of collective attacks as~\cite{mayers2001unconditional, shor2000simple, devetak2005devetak},
\begin{equation}
R = H(X \mid E)- H(X \mid Y), \label{eq:cond_entropy}.
\end{equation}
Here $X$ and $Y$ represent Alice and Bob's key systems respectively, $E$ represents the eavesdropper, and $H(.\mid.)$ is the conditional von Neumann entropy. Intuitively, Eq.~\eqref{eq:cond_entropy} follows from the fact that the secure key rate is equal to Eve's uncertainty about the raw key $X$ minus Bob's uncertainty. For our DPS-MDI protocol, the conditional entropy $H (X \mid E)$  can be expressed as~\cite{renner2008security}, 
		\begin{eqnarray}
		H(\tilde{X} \mid \tilde{E})=1-h(e_{\text{b}})-h(e_{\text{p}}), \label{eq:inf_cond_entropy}
		\end{eqnarray}
where $e_{\text{b}}$ is the bit error rate, and $e_{\text{p}}$ denotes the phase error rate.
	
We follow the finite-key analysis presented in~\cite{scarani2008quantum, renner2008security}, involving a generalization of von Neumann entropy, called the \emph{smooth} entropy. The objective of this smoothening of the regular entropic functions is to take into account the fluctuations arising from the finite signal size. As in the asymptotic case, Alice and Bob are assumed to share entangled pairs, which holds for our proposed scheme, as outlined in Sec.~\ref{sec:entanglement} above. The generalized form of Eq.~\eqref{eq:cond_entropy} in the finite-key regime can be expressed as~\cite{scarani2008quantum},
		\begin{eqnarray}
		r = H_{\xi}(X \mid E)-(\text{leak}_{\text{EC}}+\Delta)/n , \label{eq:key_rate}
		\end{eqnarray}
where $H_{\xi}(X \mid E)$ is the conditional smooth-min entropy, $\text{leak}_{\text{EC}}$ is the number of bits needed to be shared over a classical channel for error correction and
\begin{equation}
	\Delta = 2\,\log_{2}\frac{1}{[2(\varepsilon-\bar{\varepsilon} - \varepsilon_{\text{EC}})]}  + 7\sqrt{n\log_{2}(2/(\bar{\varepsilon}-\bar{\varepsilon}'))}. \label{eq:delta}
\end{equation}
Here, $\varepsilon_{\text{EC}}$ is the error probability, defined as the probability that Bob ends up with a wrong bit string after the error correction stage. $\bar{\varepsilon}$ and $\bar{\varepsilon}'$ are the smoothening parameters as mentioned in Lemma $2$ of~\cite{scarani2008quantum}. 	

We calculate $H_{\xi}(X \mid E)$ for our protocol using the asymptotic value of $H(X \mid E)$ and bound the phase error rate in terms of the bit error rate. We have shown in Appendix ~\ref{sec:bound} that the phase error rate of our protocol is bounded by the bit error rate as,
\begin{equation}
    e_{\text{p}}\leq e_{\text{b}}.
\end{equation}
In the finite-key regime Eq.~\eqref{eq:inf_cond_entropy} translates to,
 \begin{eqnarray}
		H_{\xi}(X \mid E)=1-h(\tilde{e}_{\text{b}})-h(\tilde{e}_{\text{p}}). \label{eq:finite_cond_entropy}
		\end{eqnarray}
	\begin{figure}[t!]
		\centerline{\includegraphics[width=0.5\textwidth,height=\textheight,keepaspectratio]{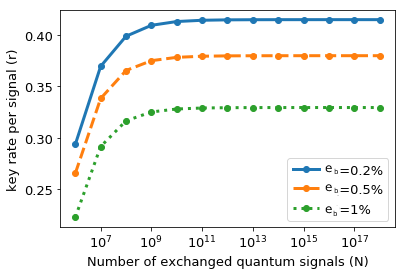}}
		\caption{Key rate $r$ as a function of the number of exchanged quantum signals for different values of $e_b$.}
		\label{fig3}
	\end{figure}
	
Finally, the bit error rate in the finite-key regime is expressed as $\tilde{e}_{\text{b}}=e_{\text{b}}+\xi \;(n,d=9)$, where $n$ is the number of raw key bits. Similarly, the phase error rate is given as $\tilde{e}_{\text{p}}=e_{\text{p}}+\xi\; (m,d=9)$, where $m$ is the number of bits used in parameter estimation and $d$ is the number of possible POVM outcomes. $d=9$ for our protocol as there are eight scenarios at the detection unit (see Table I) which contribute to the key generation. The ninth POVM corresponds to the case when BSM fails. $\xi$ is a non-negative parameter, (Lemma 3 of~\cite{scarani2008quantum}) given by,
		\begin{eqnarray}
		\xi=\sqrt{\frac{2\ln(1/\bar{\varepsilon}')+d\ln(m+1)}{m}}. \label{eq:xi}
		\end{eqnarray}
  Using Eqs.~\eqref{eq:delta},~\eqref{eq:finite_cond_entropy}, and~\eqref{eq:xi} we estimate the sifted key rate described in Eq.~\eqref{eq:key_rate}. The performance of a practical error correcting code as analyzed in \cite{scarani2008quantum} gives $\text{leak}_{\text{EC}}/n=1.2h(e_b)$, where, $e_b$ is the quantum bit error rate. This helps in estimating the second term of Eq.~\eqref{eq:key_rate}. ($N,\varepsilon,\text{leak}_{\text{EC}},\varepsilon_{\text{EC}}$) are protocol dependent parameters, whereas $n,m,\bar{\varepsilon}$ and $\bar{\varepsilon}'$ are selected so as to maximize the key rate per signal, $r=(n/N)r'$ under the constraints $n+m \leq N$ and $\varepsilon-\varepsilon_{\text{EC}} > \bar{\varepsilon} > \bar{\varepsilon}'\geq 0$.
	
Fig.~\ref{fig3} shows the variation in key rate with the number of exchanged signals for our DPS-MDI protocol. We have used $\varepsilon=10^{-5}$ and $\varepsilon_{\text{EC}}=10^{-10}$ to generate the plots for different values of $e_b$. As expected, the key rate per signal ($r$) approaches the  sifted  key  rate  of $\frac{4}{9}$ in the asymptotic  limit.  This is a reflection of the fact that only $\frac{4}{9}$ of the raw key bits can be used for key generation and the rest is used for parameter estimation.
\section{Conclusions}
In this paper, we have presented a $3$-path superposition based DPS-MDI-QKD protocol. We have shown the necessity and advantages of having the 3-path superposition. The proposed protocol has been mapped to an entanglement-based protocol, thereby establishing its unconditional security. We have carried out a security analysis of our scheme in the asymptotic regime assuming system imperfections.

We have shown that our protocol generates secure keys even when the ideal single-photon source is replaced with a weak coherent source (WCS). The security of the WCS-based scheme is established using decoy states and a suitable phase-post-selection technique. Finally, we have determined an upper-bound for the phase error rate of our protocol in terms of the bit error rate. This allows us to carry out the key analysis of the protocol in both asymptotic as well as finite-key regimes. We have further simulated the variation in key rate with the number of exchanged signals of our protocol.
	
An interesting direction for future work is the finite-key analysis of the 3-path DPS-MDI using a weak coherent source. Such a coherent-state DPS-MDI protocol will also be free from the issues arising due to the probabilistic nature of photon generation in single-photon sources. Another interesting problem that can be addressed in the future works is the tightening of the bound used in obtaining the secure key rates of our protocol.

\begin{acknowledgements}
We acknowledge the financial support by MHRD through MHRD Sanction No. F.NO. 35-8/2017-TS.1 under the Uchchatar Avishkar Yojana, with partial support from QNu Labs.	
\end{acknowledgements}

\section*{Conflict of interest}
The authors declare that they have no conflict of interest.

\bibliographystyle{apsrev4-1}
\bibliography{dps-mdi_v2.bib}

	\appendix
	\begin{widetext}
	           
		\section{Analysis of DPS-MDI-QKD protocol}\label{sec:BMS}
		\begin{center}
			\begin{figure}[htbp]
				\centerline{\includegraphics[width=0.2\textwidth,height=\textheight,keepaspectratio]{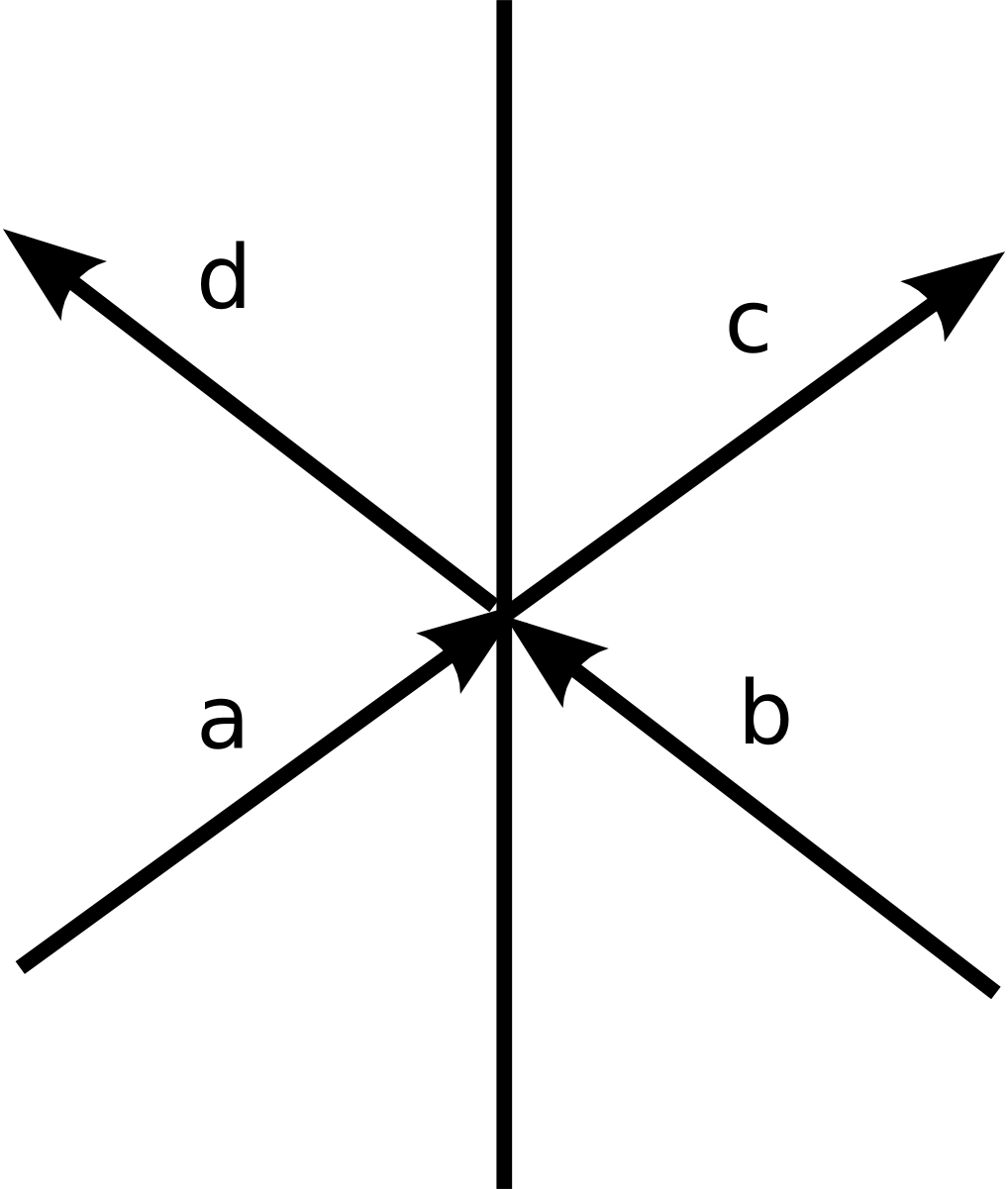}}
				\caption{$a$ and $b$ are input ports, and $c$ and $d$ are the output ports of the beamsplitter.}
				\label{fig2}
			\end{figure} 
		\end{center}
		
		We start with the form of the input to Charles' beamsplitter given in Eq.~\eqref{eq:dps_mdi2}:
		\begin{eqnarray} 
				\ket{\psi}_{\text{in}} &=& \frac{1}{3}[ \; \ket{100,100}_{ab} + e^{i\phi_{{a}_{1}}}\ket{010,100}_{ab}+ e^{i\phi_{{a}_{2}}}\ket{001,100}_{ab}+ e^{i\phi_{{b}_{1}}} \ket{100,010}_{ab} \nonumber \\
				&&
				+ e^{i\phi_{{b}_{2}}}\ket{100,001}_{ab}+ e^{i(\phi_{{a}_{1}}+\phi_{{b}_{1}})}\ket{010,010}_{ab} + e^{i(\phi_{{a}_{1}}+\phi_{{b}_{2}})} \ket{010,001}_{ab} \nonumber \\
				&&
				+ e^{i(\phi_{{a}_{2}}+\phi_{{b}_{1}})} \ket{001,010}_{ab}+ e^{i(\phi_{{a}_{2}}+\phi_{{b}_{2}})} \ket{001,001}_{ab} \; ].\nonumber
				\end{eqnarray}
		
		We leave out the states that correspond to photons traversing identical paths in Alice's and Bob's set-up, since they do not contribute to the sifted key, and consider the (normalized) state,
		\begin{eqnarray} 
						\ket{\psi}_{\text{in}} = \frac{1}{\sqrt{6}}[&& \; e^{i\phi_{{a}_{1}}}\ket{010,100}_{ab}+ e^{i\phi_{{a}_{2}}}\ket{001,100}_{ab}+ e^{i\phi_{{b}_{1}}} \ket{100,010}_{ab} + e^{i\phi_{{b}_{2}}}\ket{100,001}_{ab}  \nonumber \\
						&&  
						+ e^{i(\phi_{{a}_{1}}+\phi_{{b}_{2}})} \ket{010,001}_{ab}+ e^{i(\phi_{{a}_{2}}+\phi_{{b}_{1}})} \ket{001,010}_{ab} \; ].\label{eq:dps_mdi3}
						\end{eqnarray}
		Writing $\phi_{a_{1}}-\phi_{b_{1}}=\Delta\phi_{1}$ and $\phi_{a_{2}}-\phi_{b_{2}}=\Delta\phi_{2}$ as the phase differences between corresponding pulses from Alice and Bob, the input to Charles' beamsplitter is written as,  
		\begin{eqnarray}\label{eq:dps_mdi4}
		\ket{\tilde{\psi}}_{\text{in}} &=& \frac{1}{\sqrt{6}} e^{i(\phi_{{b}_{1}}+\phi_{{b}_{2}} )}  [\, e^{i\Delta\phi_{1}} \ket{010,001}_{ab}+  e^{i\Delta\phi_{2}}\ket{001,010}_{ab}+ e^{-i\phi_{{b}_{1}}} \Big( \ket{100,001}   \nonumber \\
		&& 
	+ e^{i\Delta\phi_{2}} \ket{001,100} \Big)	+ e^{-i\phi_{{b}_{2}}} \Big( \, \ket{100,010}   + e^{i\Delta\phi_{1}}\ket{010,100}_{ab} \Big) \, ].
		\end{eqnarray}
		
		Fig.~\ref{fig2} shows a typical $50:50$  beamsplitter. The action of the beamsplitter with input ports $a,b$ and output ports $c,d$, when there is a photon incident on only one of the two ports is given by,
	\begin{eqnarray}
	\ket{1,0}_{ab} &\longrightarrow& \frac{1}{\sqrt{2}}\left( \ket{1,0}_{cd} + \ket{0,1}_{cd} \right ), \nonumber \\
	\ket{0,1}_{ab} &\longrightarrow& \frac{1}{\sqrt{2}}\left( \ket{1,0}_{cd} - \ket{0,1}_{cd} \right ) . \label{eq:optical_bsm}
	\end{eqnarray} 
	Using Eq.~\ref{eq:optical_bsm}, we find that the beamsplitter transforms the terms present in the joint input state of Alice and Bob (Eq.~\ref{eq:dps_mdi4}) as shown below,
		
		\begin{eqnarray}
	\ket{010,001}_{ab} &\longrightarrow& \frac{1}{2}\Big(\ket{011,000}_{cd}-\ket{010,001}_{cd}+\ket{001,010}_{cd}-\ket{000,011}_{cd}\Big) , \nonumber \\
	\ket{001,010}_{ab} &\longrightarrow&\frac{1}{2} \Big(\ket{011,000}_{cd}+\ket{010,001}_{cd} -\ket{001,010}_{cd}-\ket{000,011}_{cd}\Big) , \nonumber \\
\ket{100,001}_{ab} &\longrightarrow&\frac{1}{2} \Big(\ket{101,000}_{cd}-\ket{100,001}_{cd}+\ket{001,100}_{cd}-\ket{000,101}_{cd}\Big) , \nonumber \\
\ket{001,100}_{ab} &\longrightarrow&\frac{1}{2} \Big(\ket{101,000}_{cd}+\ket{100,001}_{cd}-\ket{001,100}_{cd}-\ket{000,101}_{cd}\Big) ,\nonumber \\
	\ket{100,010}_{ab} &\longrightarrow&\frac{1}{2} \Big(\ket{110,000}_{cd}-\ket{100,010}_{cd}+\ket{010,100}_{cd}\,-\ket{000,110}_{cd}\Big) , \nonumber \\
	\ket{010,100}_{ab} &\longrightarrow&\frac{1}{2} \Big(\ket{110,000}_{cd}+\ket{100,010}_{cd} -\ket{010,100}_{cd}-\ket{000,110}_{cd}\Big) , \nonumber \\
	\ket{100,100}_{ab} &\longrightarrow&\frac{1}{\sqrt{2}} \Big(\ket{200,000}_{cd}-\ket{000,200}_{cd}\Big) , \nonumber \\
	\ket{010,010}_{ab} &\longrightarrow&\frac{1}{\sqrt{2}} \Big(\ket{020,000}_{cd}-\ket{000,020}_{cd}\Big) , \nonumber \\
	\ket{001,001}_{ab} &\longrightarrow&\frac{1}{\sqrt{2}} \Big(\ket{002,000}_{cd}-\ket{000,002}_{cd}\Big) .
	 \label{eq:bsm2}
		\end{eqnarray}
		
		Using Eq.~\eqref{eq:dps_mdi4} and Eq.~\eqref{eq:bsm2}, we get
		
		\begin{eqnarray} \label{eq:bsm_output}
		\ket{\psi}_{\text{out}}&=&\frac{1}{2\sqrt{6}}e^{i(\phi_{{b}_{1}}+\phi_{{b}_{2}})}\Big[ \, e^{i\Delta\phi_{1}}\Big(\ket{011,000}_{cd}-\ket{010,001}_{cd}+\ket{001,010}_{cd}-\ket{000,011}_{cd}\Big)\nonumber\\
		&&
		+e^{i\Delta\phi_{2}}\Big(\ket{011,000}_{cd}+\ket{010,001}_{cd} -\ket{001,010}_{cd}-\ket{000,011}_{cd}\Big)\nonumber \\
		&&
		+ e^{-i\phi_{{b}_{1}}}\Big\{\Big(\ket{101,000}_{cd}-\ket{100,001}_{cd}+\ket{001,100}_{cd}
		\,-\ket{000,101}_{cd}\Big)\nonumber \\
		&&
		+e^{i\Delta\phi_{2}}\Big(\ket{101,000}_{cd}+\ket{100,001}_{cd}-\ket{001,100}_{cd}-\ket{000,101}_{cd}\Big)\Big\}\nonumber \\
		&&
		+ e^{-i\phi_{{b}_{2}}}\Big\{\Big(\ket{110,000}_{cd}
		\,-\ket{100,010}_{cd}+\ket{010,100}_{cd}\,-\ket{000,110}_{cd}\Big)\nonumber \\
		&&
		+e^{i\Delta\phi_{1}}\Big(\ket{110,000}_{cd}+\ket{100,010}_{cd} -\ket{010,100}_{cd}-\ket{000,110}_{cd}\Big)\Big\}\Big].
		\end{eqnarray}

The output after the beamsplitter depends upon the random phase applied by Alice and Bob to their respective time-bins. We write down the four different final states realized, corresponding to the four possible values of $(\Delta\phi_{1}, \Delta\phi_{2})$. To help understand the key-reconciliation step, we have rewritten the final state by grouping together the states at each output port ($c$ or $d$), corresponding to the three different time-bins ($t_{1}$, $t_{2}$ or $t_{3}$). \\
		
		\textbf{Case 1:} When $\Delta\phi_{1}=\Delta\phi_{2}=0$, the two-photon state after the beamsplitter is,
		\begin{eqnarray}\label{eq:dps_mdi7}
		\ket{\psi}_{\text{out}}=\frac{1}{\sqrt{6}}e^{i(\phi_{{b}_{1}}+\phi_{{b}_{2}})}\Big[&& \Big(\ket{011,000}_{cd}-\ket{000,011}_{cd}\Big)		+e^{-i\phi_{{b}_{1}}}\Big(\ket{101,000}_{cd}\nonumber \\
		&&
		-\ket{000,101}_{cd}\Big)+e^{-i\phi_{{b}_{2}}}\Big(\ket{110,000}_{cd}-\ket{000,110}_{cd}\Big)\Big].
		\end{eqnarray}
		
		\textbf{Case 2:} When $\Delta\phi_{1}=\Delta\phi_{2}=\pi$, the output state of the beamsplitter is,
		\begin{eqnarray}\label{eq:dps_mdi8}
		\ket{\psi}_{\text{out}}=\frac{1}{\sqrt{6}}e^{i(\phi_{{b}_{1}}+\phi_{{b}_{2}})}\Big[&& \Big(\ket{011,000}_{cd}-\ket{000,011}_{cd}
				+e^{-i\phi_{{b}_{1}}}\Big(\ket{001,100}_{cd}\nonumber \\
		&&
		-\ket{100,001}_{cd}\Big)+e^{-i\phi_{{b}_{2}}}\Big(\ket{010,100}_{d}-\ket{100,010}_{d}\Big)\Big].
		\end{eqnarray}
		
		\textbf{Case 3:} When $\Delta\phi_{1}=0$ and $\Delta\phi_{2}=\pi$, the output state is,
		\begin{eqnarray}\label{eq:dps_mdi9}
		\ket{\psi}_{\text{out}}=\frac{1}{\sqrt{6}}e^{i(\phi_{{b}_{1}}+\phi_{{b}_{2}})}\Big[&& \Big(\ket{001,010}_{cd}-\ket{010,001}_{cd}\Big)+e^{-i\phi_{{b}_{1}}}\Big(\ket{001,100}_{cd}\nonumber \\		
		&&
		-\ket{100,001}_{cd}\Big)+e^{-i\phi_{{b}_{2}}}\Big(\ket{110,000}_{cd}-\ket{000,110}_{cd}\Big)\Big].
		\end{eqnarray}
		
		\textbf{Case 4:} When  $\Delta\phi_{1}=\pi$ and $\Delta\phi_{2}=0$, the output state is,
		\begin{eqnarray}\label{eq:dps_mdi10}
		\ket{\psi}_{\text{out}}=\frac{1}{\sqrt{6}}e^{i(\phi_{{b}_{1}}+\phi_{{b}_{2}})}\Big[&& \Big(\ket{010,001}_{cd}-\ket{001,010}_{cd}\Big)+e^{-i\phi_{{b}_{1}}}\Big(\ket{101,000}_{cd}\nonumber \\
		&&
		-\ket{000,101}_{cd}\Big)+e^{-i\phi_{{b}_{2}}}\Big(\ket{010,100}_{cd}-\ket{100,010}_{cd}\Big)\Big].
		\end{eqnarray}

 We now formulate the key reconciliation scheme (see Table I) based on Eqs.~\eqref{eq:dps_mdi7}-~\eqref{eq:dps_mdi10}, while noting that detector $D_{c}$ detects the photons from port $c$ of the beamsplitter and correspondingly detector $D_{d}$ clicks when photons exits from port $d$. 
 
 Consider the two examples when Alice and Bob use $\Delta\phi_{1}$ to extract the key.
 \begin{enumerate}
 \item  When Charles announces the clicking of $D_c$ in time-bins $t_{1}$ and $t_{2}$, this would indicate that $\Delta\phi_{1}$ and $\Delta\phi_{2}$ have taken values corresponding to Case 1 or Case 3 above, corresponding to $\Delta\phi_{1} = 0$ and $\Delta\phi_{2}= 0$ or $\pi$. Alice and Bob therefore use only $\Delta\phi_{1}$ to extract the key.
 \item When Charles announces the clicking of $D_c$ at $t_{1}$ and $D_d$ at $t_{2}$, Alice and Bob again use $\Delta\phi_{1}$ to extract the key. However, they also need a bit flip operation to get the same key bits. Note that in this example also, $\Delta\phi_{1}=0$ and $\Delta\phi_{2}=0$ or $\pi$.
 \end{enumerate}
A similar reasoning can be used to complete the key reconciliation scheme as described in Table 1.
	\section{DPS-MDI as an entanglement-based protocol}\label{sec:EB}
We start with Eqs.~\eqref{eq:alice} and~\eqref{eq:Bob_encoded_state}, to write the joint state of Alice and Bob after their encoding procedure. Recall that $A$ and $B$ indicate Alice and Bob's signal states, whereas $A_{i}$ and $B_{i}$ indicate the $i^{\rm th}$ pair of ancilla qubit in respective (ideal) quantum memories. The joint state thus reads as,
\begin{eqnarray}
\ket{\psi}_{\text{Alice}}\otimes |\psi\rangle_{\text{Bob}} &=& \frac{1}{4}\sum_{j_{1},j_{2}\in \{ 0,1 \}}(\ket{j_{1}}_{A_{1}}\ket{j_{2}}_{A_{2}})\otimes\ket{\psi_{j_1j_2}}_{a}\otimes \sum_{\tilde{j}_{1},\tilde{j}_{2} \in \{ 0,1 \}}(\ket{\tilde{j}_{1}}_{B_{1}}\ket{\tilde{j}_{2}}_{B_{2}})\otimes\ket{\psi_{\tilde{j}_{1},\tilde{j}_{2}}}_{b}, \nonumber \\
&=& \frac{1}{4}\sum_{j_{1},j_{2},\tilde{j}_{1},\tilde{j}_{2} \in \{ 0,1 \}}  \ket{j_{1}}_{A_{1}}\ket{\tilde{j}_{1}}_{B_{1}}\ket{j_{2}}_{A_{2}}\ket{\tilde{j}_{2}}_{B_{2}} \otimes |\Psi_{(j_{1}j_{2}\tilde{j}_{1}\tilde{j}_{2})}\rangle_{ab}.\label{eq:input}
\end{eqnarray}
The state $|\Psi_{j_{1}j_{2}\tilde{j}_{1}\tilde{j}_{2}}\rangle_{ab}$, which eventually becomes the input to Charles' beamsplitter, has the following form:
\begin{eqnarray}
\vert \Psi_{j_{1}j_{2}\tilde{j}_{1}\tilde{j}_{2}}\big\rangle_{ab} &=&\ket{\psi_{j_1j_2}}_{a}\otimes \ket{\psi_{\tilde{j_1}\tilde{j_2}}}_{b}\nonumber \\
&=&  
 \Big( a^{\dagger}_{1}b_{1}^{\dagger} + \sum_{i=1}^{2}(-1)^{(j_{i}+\tilde{j}_{i})}a_{i+1}^{\dagger}b_{i+1}^{\dagger} + (-1)^{\tilde{j}_{1}}a_{1}^{\dagger}b_{2}^{\dagger} + (-1)^{ \tilde{j}_{2}}a_{1}^{\dagger}b_{3}^{\dagger} \nonumber \\
&+& (-1)^{j_{1}}a_{2}^{\dagger}b_{1}^{\dagger} + (-1)^{j_{1}+ \tilde{j}_{2}}a_{2}^{\dagger}b_{3}^{\dagger} + (-1)^{j_{2}} a_{3}^{\dagger}b_{1}^{\dagger} + (-1)^{(j_{2} + \tilde{j}_{1})}a_{3}^{\dagger}b_{2}^{\dagger} \Big) \vert 0,0 \big\rangle_{ab} . \label{eq:dps-mdi-ent}
\end{eqnarray}
Here, $\ket{0,0}_{ab}=\ket{0}_a\otimes\ket{0}_b$, and denotes the vaccum at the input ports of the beamsplitter.  $a_{i}^{\dagger}$ and $b_{i}^{\dagger}$ are the creation operators corresponding to a photon traversing through the $i^{\rm th}$ arm in Alice and Bob's delay lines respectively. As indicated above, there is no entanglement yet between Alice and Bob's states; rather, each encoded state is entangled with their respective quantum memories. 

To obtain the output state after measurement and key reconciliation, we first do a post-selection and discard input states which have photons arriving at the same time-bin from both Alice and Bob. As described in section \ref{sec:dps-mdi}, such photons do not contribute to the final key, due to Hong-Ou-Mandel interference. Hence, we drop terms of the form $a_{i}^{\dagger}b_{i}^{\dagger}$ in Eq.~\eqref{eq:dps-mdi-ent}. When the photons arrive at different times, as represented by terms of the form $a_{i}^{\dagger}b_{j}^{\dagger}$ for $i\neq j$, they transform as,
\[ a^{\dagger} \rightarrow \frac{1}{\sqrt{2}}(c^{\dagger} + d^{\dagger})\, ; \quad b^{\dagger} \rightarrow \frac{1}{\sqrt{2}}(c^{\dagger} -d^{\dagger}) .\]

We may thus write down the final state after the action of the beamsplitter and post-selection as,
\begin{eqnarray}
 \vert \Phi_{j_{1}j_{2}\tilde{j}_{1}\tilde{j}_{2}}\big\rangle_{cd} &=& 
 \frac{1}{2}\Big[ (-1)^{\tilde{j}_{1}}(c_{1}^{\dagger}+ d_{1}^{\dagger})(c_{2}^{\dagger} - d_{2}^{\dagger}) + (-1)^{ \tilde{j}_{2}}(c_{1}^{\dagger}+ d_{1}^{\dagger})(c_{3}^{\dagger} - d_{3}^{\dagger}) + (-1)^{j_{1}}(c_{2}^{\dagger}+ d_{2}^{\dagger})\times \nonumber \\
 && (c_{1}^{\dagger} - d_{1}^{\dagger}) +  (-1)^{(j_{1} + \tilde{j}_{2})}(c_{2}^{\dagger}+ d_{2}^{\dagger})(c_{3}^{\dagger} - d_{3}^{\dagger}) + (-1)^{j_{2}} (c_{3}^{\dagger} + d_{3}^{\dagger})(c_{1}^{\dagger} - d_{1}^{\dagger}) \nonumber \\ 
 && +  (-1)^{(j_{2} + \tilde{j}_{1})}(c_{3}^{\dagger} + d_{3}^{\dagger})(c_{2}^{\dagger} - d_{2}^{\dagger}) \Big]\vert 0,0\rangle_{cd} . \label{eq:dps-mdi-ent2}
\end{eqnarray}
The complete state, including the registers $A_{1}, A_{2}$ and $B_{1}, B_{2}$, is of the form,
\begin{equation}
|\chi\rangle_{A_{1}B_{1}A_{2}B_{2}cd} = \frac{1}{4}\sum_{j_{1},j_{2},\tilde{j}_{1},\tilde{j}_{2} \in \{ 0,1 \}}  \ket{j_{1}}_{A_{1}}\ket{\tilde{j}_{1}}_{B_{1}}\ket{j_{2}}_{A_{2}}\ket{\tilde{j}_{2}}_{B_{2}} \otimes |\Phi_{(j_{1}j_{2}\tilde{j}_{1}\tilde{j}_{2})}\rangle_{cd}. \label{eq:dps-mdi-ent3}
\end{equation}
	
As discussed in section ~\ref{sec:entanglement} , Alice and Bob extract information about their relative phases $\Delta \phi_{1} = \phi_{a_{1}}-\phi_{b_{1}}$  and $\Delta \phi_{2} = \phi_{a_{2}}-\phi_{b_{2}}$ based on Charles' measurement outcomes, and hence obtain the shared key. Expressing all the phases in terms of the binary variables $(j_{1}, j_{2})$ and $(\tilde{j}_{1}, \tilde{j}_{2})$, which characterize Alice and Bob's qubit registers respectively, we have, 
\[ \phi_{a_{1}} = j_{1}\pi, \; \phi_{a_{2}} = j_{2} \pi, \; \phi_{b_{1}} = \tilde{j}_{1}\pi, \; \phi_{b_{2}} = \tilde{j}_{2} \pi . \]
Thus the relative phases are given by,
\[ \Delta\phi_{1} = (j_{1} - \tilde{j}_{1})\pi, \; \; \Delta\phi_{2} = ( j_{2} - \tilde{j}_{2} )\pi .\]

It is now easy to show that the joint state of Alice and Bob's registers collapses to an entangled state after Charles' measurement and the reconciliation process described in Table $1$. In particular, when Alice and Bob use the phases $\phi_{a_{i}}, \phi_{b_{i}}$ to generate their secret key bits without a bit flip operation, they end up with the perfectly correlated Bell state $\frac{1}{\sqrt{2}}[\ket{00}_{A_{i}B_{i}} -\ket{11}_{A_{i}B_{i}}]$. In those cases where they need to perform a bit flip operation, they end up sharing the anti-correlated entangled state $\frac{1}{\sqrt{2}}[\ket{10}_{A_{i}B_{i}} -\ket{01}_{A_{i}B_{i}}]$. 

For example, 
when Charles announces that the detector $c$ has clicked in both $t_{1}$ and $t_{2}$ bins, Eq.~\eqref{eq:dps-mdi-ent3} collapses to the post-measurement state,
\begin{eqnarray}\label{eq:dps-mdi-ent4}
|\chi^{(1)}\rangle_{\rm out} &=& \frac{1}{2\sqrt{2}} \Big[ \ket{0000}_{A_{1}B_{1}A_{2}B_{2}} - \ket{0001}_{A_{1}B_{1}A_{2}B_{2}} + \ket{0010}_{A_{1}B_{1}A_{2}B_{2}} - \ket{0011}_{A_{1}B_{1}A_{2}B_{2}} \nonumber \\
	&& 		- \ket{1100}_{A_{1}B_{1}A_{2}B_{2}} + \ket{1101}_{A_{1}B_{1}A_{2}B_{2}} - \ket{1110}_{A_{1}B_{1}A_{2}B_{2}} +\ket{1111}_{A_{1}B_{1}A_{2}B_{2}} \Big]\nonumber \\
	&&
	\otimes \ket{110,000}_{cd},
\end{eqnarray}
where we have represented $\ket{j_1}_{A_1}\ket{\tilde{j_1}}_{B_1}\ket{j_2}_{A_2}\ket{\tilde{j_2}}_{B_2}$ as $\ket{j_1\tilde{j}_1j_2\tilde{j}_2}_{A_{1}B_{1}A_{2}B_{2}}$. We see that in Eq.~\eqref{eq:dps-mdi-ent4}, the first ancilla registers ($A_{1}$ and $B_{1}$) of both Alice and Bob always have same bit value. 
Hence, Alice and Bob share the perfectly correlated Bell state, as shown explicitly below,
\begin{eqnarray}\label{eq:dps-mdi-ent5}
\ket{\chi^{(1)}}_{\rm out} &=& \frac{1}{2\sqrt{2}} \Big[\ket{00}_{A_{1}B_{1}} - \ket{11}_{A_{1}B_{1}} \Big] \otimes \Big[ \ket{00}_{A_{2}B_{2}} - \ket{01}_{A_{2}B_{2}} + \ket{10}_{A_{2}B_{2}} - \ket{11}_{A_{2}B_{2}} \Big]\nonumber \\
&&
\otimes \ket{110,000}_{cd}. 
\end{eqnarray}

When Charles announces that the detector $c$ clicked at $t_{1}$ and $d$ at $t_{2}$, the state presented in Eq.~\eqref{eq:dps-mdi-ent3} collapses to,
\begin{eqnarray}\label{eq:dps-mdi-ent6}
\ket{\chi^{(2)}}_{\rm out}&=&\frac{1}{2\sqrt{2}}\Big[-\ket{0100}_{A_{1}B_{1}A_{2}B_{2}}+\ket{0101}_{A_{1}B_{1}A_{2}B_{2}} - \ket{0110}_{A_{1}B_{1}A_{2}B_{2}} + \ket{0111}_{A_{1}B_{1}A_{2}B_{2}} \nonumber\\
		&&
		-\ket{1000}_{A_{1}B_{1}A_{2}B_{2}} -\ket{1001}_{A_{1}B_{1}A_{2}B_{2}} + \ket{1010}_{A_{1}B_{1}A_{2}B_{2}} - \ket{1011}_{A_{1}B_{1}A_{2}B_{2}} \Big]\nonumber \\
		&&
		\otimes  \ket{100,010}_{cd} .
		\end{eqnarray}
As seen from Eq.~\eqref{eq:dps-mdi-ent6}, the first ancilla registers ($A_{1}$ and $B_{1}$) of Alice and Bob are now always opposite in the bit value. This implies they share an anti-correlated entangled state. Hence, they require a bit flip operation after Charles announcement so as to ensure that both of them end up with similar key bits. We can extend similar lines of reasoning to the other entries of Table $1$ to show that Alice and Bob indeed share maximally entangled states.
\section{Bounding of phase error rate in terms of bit error rate}\label{sec:bound}
In Appendix ~\ref{sec:EB}, we show that Charles measurement entangles Alice's and Bob's ancilla qubits. However, the EPR pairs shared by Alice and Bob become corrupt due to channel noise and eavesdropping. Alice and Bob extract a small number of perfect EPR pairs from the corrupted EPR pairs using a suitable entanglement distillation protocol based on Calderbank-Shor-Steane (CSS) codes, provided the channel is not too noisy ~\cite{shor2000simple}. Alice and Bob determine the bit and the phase error rates. They continue with the entanglement distillation protocol if the error rates are nominal, else they abort the protocol. The bit error rates can be easily estimated by sharing a certain fraction of the raw key generated during the experiment. However, phase errors cannot be determined experimentally, and hence, need to be estimated indirectly using experimentally observed quantities. We upper bound the phase error rate for our scheme in terms of the bit error rate in this section. 

We begin with $\ket{\psi}^{(l)}_{\text{out}}$, which is the state after Charles announces his measurement result for the $l^{\text{th}}$ time slot, and is related to the joint input state of Alice and Bob $\ket{\psi}^{(l)}_{\text{in}}$ as, 
\begin{equation}
\ket{\psi}_{\text{out}}^{(l)}=F^{(l)}M^{(l)}E^{(l)}\ket{\psi}^{(l)}_{\text{in}}.\label{eq:out}
\end{equation}	
 Here, $M^{(l)}$ is the beamsplitter operator acting on the $l^{\text{th}}$ time slot, $F^{(l)}$ is the filtering operator and $E^{(l)}$ is a $3 \times 3$``noise" matrix representing the effects of noise and Eve's most general attack in $l^{\text{th}}$ time slot. We assume that the noise and Eve affect the link connecting Alice to Charles and Bob to Charles independently. Hence, we decompose the overall noise matrix $E^{(l)}$ as $E_{a}^{(l)}\otimes E_b^{(l)}$. Both $E_a^{(l)}$ and $E_b^{(l)}$ are  $3\times 3$ matrices with matrix elements $(a)_{ij}$ and $(b)_{ij}$ respectively. $\vert a_{ij}\vert^2$ gives the probability of time-bin $i$ getting affected given that the noise/Eve acts on time-bin $j$. We would like to clarify the terms ``time-bin" and ``time-slot" used here. We use the time-slot label to mark every single-photon state (in the ideal scenario) or weak coherent pulse (in a typical experiment) generated by the source, whether Alice or Bob. Each pulse labelled by a time-slot is eventually measured at one of three time-bins ($i/j =1, 2, 3$) by Charlie, depending on which path the photon traversed in the DLI at the source. The form of these matrices depends upon the type of  noise in the channel. The matrix structure is dependent upon the eavesdropper's attack too. Hereafter, in the interest of brevity, we drop the superscript (l).
 
 For example, we can study a channel noise (or an attack) which flips  the key bits. The key is encoded in the phase difference of the corresponding time-bins of Alice and Bob in our protocol. Hence, Eve would need to flip the phase of Alice's time-bins or Bob's time-bins. So different noise matrices that can lead to such an attack are
 \begin{equation}
\begin{pmatrix}
 1 & 0 & 0\\
 0 & -1 & 0\\
 0 & 0 & -1
 \end{pmatrix}_{a}\otimes
 \begin{pmatrix}
  1 & 0 & 0\\
  0 & 1 & 0\\
  0 & 0 & 1
  \end{pmatrix}_{b}\quad \text{or} \quad  
 \begin{pmatrix}
    1 & 0 & 0\\
    0 & 1 & 0\\
    0 & 0 & 1
    \end{pmatrix}_{a}\otimes
 \begin{pmatrix}
 1 & 0 & 0\\
 0 & -1 & 0\\
 0 & 0 & -1
 \end{pmatrix}_{b}     
 \end{equation}
Another example is of an attack where Eve just monitors the presence of a photon in the second time-bin of Alice's signal. We can write Alice's state as 
\begin{equation}
\ket{\psi }_{\text{in}}=\frac{1}{\sqrt{3}} \left( \, \ket{100}_{a} + e^{i\phi_{{a}_{1}}} \ket{011}_{a} + e^{i\phi_{{a}_{2}}} \ket{001}_{a} \, \right)\label{eq:attack}
\end{equation}
When Eve discovers no photon in the second time-bin, the state shown in Eq.~\eqref{eq:attack} collapses to 
\begin{equation}
\ket{\psi }_{\text{final}}=\frac{1}{\sqrt{2}} \left( \, \ket{100}_{a}+ e^{i\phi_{{a}_{2}}} \ket{001}_{a} \, \right)
\end{equation}
One such noise matrix  that achieves this attack is
 \begin{equation}
 E_a=\begin{pmatrix}
    \frac{1}{\sqrt{6}} & \frac{1}{\sqrt{6}} & \frac{1}{\sqrt{6}}\\
    0 & 0 & 0\\
    \frac{1}{\sqrt{6}} & \frac{1}{\sqrt{6}} & \frac{1}{\sqrt{6}}
    \end{pmatrix}
\end{equation}

From the above examples we conclude that the elements of these noise matrices can be predicted only when we know the nature of Eve's attack and the noise in the channel. Hence, by assuming a general form for these matrices, we can find the bit and phase error rates in our protocol for any general eavesdropping strategy.

Alice and Bob measure their EPR pairs in the $Z$ ($X$) basis, which acts as a stabilizer for the bit (phase) error. Thus, the probability of obtaining a bit error in the $l^{\text{th}}$ time-slot is,
	\begin{equation} 
		e{_{\text{b}}}=1-\frac{1}{2}\big(\langle\psi\vert I_{cd}\otimes Z_{A_{1}B_{1}}\otimes I_{A_{2}B_{2}}\vert\psi_{\text{out}}\rangle + \langle\psi_{\text{out}}\vert I_{cd}\otimes I_{A_{1}B_{1}}\otimes Z_{A_{2}B_{2}}\vert\psi_{\text{out}}\rangle\big)\label{eq:ber}.
		\end{equation} 	
Similarly, the probability of obtaining a phase error in the $l^{\text{th}}$ time-slot can be expressed as,
	\begin{equation} 
		e_{\text{p}}=1-\frac{1}{2}\big(\langle\psi_{\text{out}}\vert I_{cd}\otimes X_{A_{1}B_{1}}\otimes I_{A_{2}B_{2}}\vert\psi_{\text{out}}\rangle + \langle\psi_{\text{out}}\vert I_{cd}\otimes I_{A_{1}B_{1}}\otimes X_{A_{2}B_{2}}\vert\psi_{\text{out}}\rangle\big)\label{eq:per}.
		\end{equation}
Using Eq.~\eqref{eq:out}, we re-write the bit error rate as,
\begin{eqnarray}
e{_{\text{b}}}&=&1-\frac{1}{2}\big(\langle\psi_{\text{in}}\vert E^{\dagger}M^{\dagger}F^{\dagger}_1 F_1M E\otimes Z_{A_{1}B_{1}}\otimes I_{A_{2}B_{2}}\vert\psi_{\text{in}}\rangle \nonumber \\ 
&&
+ \langle\psi_{\text{in}}\vert E^{\dagger}M^{\dagger}F^{\dagger}_2 F_2ME\otimes I_{A_{1}B_{1}}\otimes Z_{A_{2}B_{2}}\vert\psi_{\text{in}}\rangle\big).\label{eq:qber_A}
\end{eqnarray}
Here, $F_{1}$ and $F_{2}$ are the filtering operators corresponding to the instances where Charles measurement and its public announcement effectively results in entangling the first and second ancilla qubits of Alice and Bob, respectively.
\subsection{Decomposing $M^{(l)\dagger}F^{(l)\dagger}_1 F^{(l)}_1M^{(l)}$ }
$\ket{100}$, $\ket{010}$ and $\ket{001}$ form an orthonormal basis for Alice's and Bob's system, where e.g. $\ket{100}$ represents a photon in the first time-bin. Using Eq.~\eqref{eq:bsm2}, for each time slot ($l$), we write,
\begin{eqnarray}
M&=&\frac{1}{2}\Big[\big(\ket{100}_c\bra{100}_a+\ket{100}_d\bra{100}_a\big)\otimes \big(\ket{100}_c\bra{100}_b-\ket{100}_d\bra{100}_b\big)\nonumber \\
&&
+\big(\ket{100}_c\bra{100}_a+\ket{100}_d\bra{100}_a\big)\otimes \big(\ket{010}_c\bra{010}_b-\ket{010}_d\bra{010}_b\big)\nonumber \\
&&
+\big(\ket{100}_c\bra{100}_a+\ket{100}_d\bra{100}_a\big)\otimes \big(\ket{001}_c\bra{001}_b-\ket{001}_d\bra{001}_b\big)\nonumber \\
&&
+\big(\ket{010}_c\bra{010}_a+\ket{010}_d\bra{010}_a\big)\otimes \big(\ket{100}_c\bra{100}_b-\ket{100}_d\bra{100}_b\big)\nonumber \\
&&
+\big(\ket{010}_c\bra{010}_a+\ket{010}_d\bra{010}_a\big)\otimes \big(\ket{010}_c\bra{010}_b-\ket{010}_d\bra{010}_b\big)\nonumber \\
&&
+\big(\ket{010}_c\bra{010}_a+\ket{010}_d\bra{010}_a\big)\otimes \big(\ket{001}_c\bra{001}_b-\ket{001}_d\bra{001}_b\big)\nonumber \\
&&
+\big(\ket{001}_c\bra{001}_a+\ket{001}_d\bra{001}_a\big)\otimes \big(\ket{100}_c\bra{100}_b-\ket{100}_d\bra{100}_b\big)\nonumber \\
&&
+\big(\ket{001}_c\bra{001}_a+\ket{001}_d\bra{001}_a\big)\otimes \big(\ket{010}_c\bra{010}_b-\ket{010}_d\bra{010}_b\big)\nonumber \\
&&
+\big(\ket{001}_c\bra{001}_a+\ket{001}_d\bra{001}_a\big)\otimes \big(\ket{001}_c\bra{001}_b-\ket{001}_d\bra{001}_b\big)\Big].\nonumber \\ \label{eq:BS_unitary}
\end{eqnarray}
$F_1$ acts as identity for the measurement results which contribute towards the key. Hence, it can be expressed as,
\begin{eqnarray}
F_1&=&\ket{110}_c\bra{110}_c\otimes\ket{000}_d\bra{000}_d+\ket{000}_c\bra{000}_c\otimes\ket{110}_d\bra{110}_d\nonumber \\
&&
\ket{100}_c\bra{100}_c\otimes\ket{010}_d\bra{010}_d+\ket{010}_c\bra{010}_c\otimes\ket{100}_d\bra{100}_d. \label{eq:filter}
\end{eqnarray}
Using Eq.~\eqref{eq:BS_unitary} and Eq.~\eqref{eq:filter}, we get
\begin{equation}
M^{\dagger}F^{\dagger}_1 F_1M=\ket{100}_a\bra{100}_a\otimes\ket{010}_b\bra{010}_b+\ket{010}_a\bra{010}_a\otimes\ket{100}_b\bra{100}_b .\label{eq:decomp}
\end{equation}
We express Eq.~\eqref{eq:decomp} in the basis of $A\otimes B$. In a concise notation, we use $\ket{a_{i}b_{i}}$ to denote the basis of the system $A\otimes B$, where e.g. $\ket{a_{1}b_{1}}$ equals $\ket{100}_a\otimes \ket{100}_b$. Using a completeness relation, we can write Eq.~\eqref{eq:decomp} as,
\begin{equation}
M^{\dagger}F^{\dagger}_1 F_1M=\ket{a_1b_2}\bra{a_1b_2}+\ket{a_2b_1}\bra{a_2b_1}.\label{eq:fil1}
\end{equation}
By defining a suitable $F_{2}$, we can write
\begin{equation}
M^{\dagger}F^{\dagger}_2 F_2M=\ket{a_1b_3}\bra{a_1b_3}+\ket{a_3b_1}\bra{a_3b_1}.\label{eq:filter2}
\end{equation}
\subsection{Bit error rate (BER)}
 First, we evaluate one component of the bit error rate - $\langle\psi_{\text{in}}\vert E^{\dagger}\ket{a_1b_2}\bra{a_1b_2}E\otimes Z_{A_{1}B_{1}}\otimes I_{A_{2}B_{2}}\vert\psi_{\text{in}}\rangle$. We assume that Eve acts independently on the channels connecting Alice to Charles and Bob to Charles. Hence, we can write $E=E_a\otimes E_b$. 
 Using Eq.~\eqref{eq:input} and Eq.~\eqref{eq:dps-mdi-ent}, we express $\langle\psi_{\text{in}}\vert E^{\dagger}\ket{a_1b_2}\bra{a_1b_2}E\otimes Z_{A_{1}B_{1}}\otimes I_{A_{2}B_{2}}\vert\psi_{\text{in}}\rangle$  as,
 \begin{eqnarray}
 \sum\limits_{\ket{j}=\ket{0000}}^{\ket{1111}}&&\Big(\bra{a_1b_1}+(-1)^{j_1+\tilde{j_1}}\bra{a_2b_2}+(-1)^{j_2+\tilde{j_2}}\bra{a_3b_3}+(-1)^{\tilde{j_1}}\bra{a_1b_2}+(-1)^{\tilde{j_2}}\bra{a_1b_3}\nonumber \\
 &&
 +(-1)^{j_1}\bra{a_2b_1}+(-1)^{j_1+\tilde{j_2}}\bra{a_2b_3}+(-1)^{j_2}\bra{a_3b_1}+(-1)^{j_2+\tilde{j_1}}\bra{a_3b_2}\Big)\nonumber \\
 &&
 \otimes \bra{j_1\tilde{j_1}j_2\tilde{j_2}} E^{\dagger}_a\otimes E^{\dagger}_b\ket{a_1b_2}\bra{a_1b_2}E_a\otimes E_b
 \otimes Z_{A_{1}B_{1}}\otimes I_{A_{2}B_{2}}\Big(\ket{a_1b_1}\nonumber \\
 &&
 +(-1)^{j_1+\tilde{j_1}}\ket{a_2b_2}+(-1)^{j_2+\tilde{j_2}}\ket{a_3b_3}+(-1)^{\tilde{j_1}}\ket{a_1b_2}+(-1)^{\tilde{j_2}}\ket{a_1b_3}\nonumber\\
 &&
 +(-1)^{j_1}\ket{a_2b_1}+(-1)^{j_1+\tilde{j_2}}\ket{a_2b_3}+(-1)^{j_2}\ket{a_3b_1}+(-1)^{j_2+\tilde{j_1}}\ket{a_3b_2}\Big)\nonumber \\
 &&
 \otimes \ket{j_1\tilde{j_1}j_2\tilde{j_2}}, \label{eq:ber1}
 \end{eqnarray}
where, $\ket{j}=\ket{j_{1}\tilde{j_1}j_2\tilde{j_2}}$ is the state of the joint quantum memory of Alice and Bob. We define the $(a_{ij})^\text{th}$ matrix elements of $E_a$ as $\bra{a_i}E_a\ket{a_j}$ and the $(b_{ij})^\text{th}$ element of $E_b$ as $\bra{b_i}E_a\ket{b_j}$. Hence, we write Eq.~\eqref{eq:ber1} as,
 \begin{eqnarray}
  \sum\limits_{\ket{j}=\ket{0000}}^{\ket{1111}}(-1)^{(j_1+\tilde{j_1})}\Big[&&\Big(a_{11}^*b_{21}^*+(-1)^{j_1+\tilde{j_1}}a_{12}^*b_{22}^*+(-1)^{j_2+\tilde{j_2}}a_{13}^*b_{23}^*+(-1)^{\tilde{j_1}}a_{11}^*b_{22}^*\nonumber \\
  &&
  +(-1)^{\tilde{j_2}}a_{11}^*b_{23}^*+(-1)^{j_1}a_{12}^*b_{21}^*+(-1)^{j_1+\tilde{j_2}}a_{12}^*b_{23}^*+(-1)^{j_2}a_{13}^*b_{21}^*\nonumber \\
  &&
  +(-1)^{j_2+\tilde{j_1}}a_{13}^*b_{22}^*\Big)\times\Big(a_{11}b_{21}+(-1)^{j_1+\tilde{j_1}}a_{12}b_{22}+(-1)^{j_2+\tilde{j_2}}a_{13}b_{23}\nonumber \\
    &&
   +(-1)^{\tilde{j_1}}a_{11}b_{22}+(-1)^{\tilde{j_2}}a_{11}b_{23} +(-1)^{j_1}a_{12}b_{21}+(-1)^{j_1+\tilde{j_2}}a_{12}b_{23}\nonumber \\
   &&
   +(-1)^{j_2}a_{13}b_{21}+(-1)^{j_2+\tilde{j_1}}a_{13}b_{22}\Big)\Big]
   \label{eq:ber3}.
 \end{eqnarray}
  Eq.~\eqref{eq:ber3} can be factorised as,
  \begin{eqnarray}
   \sum\limits_{\ket{j}=\ket{0000}}^{\ket{1111}}(-1)^{j_1+\tilde{j_1}}&&\Big(a_{11}^*+(-1)^{j_1}a_{12}^*+(-1)^{j_2}a_{13}^*\Big)\Big(b_{21}^*+(-1)^{\tilde{j_1}}b_{22}^*+(-1)^{\tilde{j_2}}b_{23}^*\Big)\nonumber \\
  &&
  \times\Big(a_{11}+(-1)^{j_1}a_{12}+(-1)^{j_2}a_{13}\Big)\Big(b_{21}+(-1)^{\tilde{j_1}}b_{22}+(-1)^{\tilde{j_2}}b_{23}\Big),\label{eq:ber4}
\end{eqnarray} 
  which further simplifies to,
 \begin{equation}
 16\Big(\vert a_{12}\vert^2+\vert a_{11}\vert^2-\vert a_{12}-a_{11}\vert^2\Big)\Big(\vert b_{22}\vert^2+\vert b_{21}\vert^2-\vert b_{22}-b_{21}\vert^2\Big).\label{eq:ber6}
 \end{equation}
 Now, we evaluate the remaining terms of Eq.~\eqref{eq:qber_A} and obtain the total BER in each time slot as,
    \begin{eqnarray}
    e_{\text{b}}=1-\frac{16}{2\times144}&&\Bigg[\Big(\vert a_{12}\vert^2+\vert a_{11}\vert^2-\vert a_{12}-a_{11}\vert^2\Big)\Big(\vert b_{22}\vert^2+\vert b_{21}\vert^2-\vert b_{22}-b_{21}\vert^2\Big)\nonumber \\
    &&
    +\Big(\vert a_{21}\vert^2+\vert a_{22}\vert^2-\vert a_{21}-a_{22}\vert^2\Big)\Big(\vert b_{12}\vert^2+\vert b_{11}\vert^2-\vert b_{12}-b_{11}\vert^2\Big)\nonumber \\
    &&
    +\Big(\vert a_{13}\vert^2+\vert a_{11}\vert^2-\vert a_{13}-a_{11}\vert^2\Big)\Big(\vert b_{33}\vert^2+\vert b_{31}\vert^2-\vert b_{33}-b_{31}\vert^2\Big)\nonumber \\
    &&
    +\Big(\vert a_{33}\vert^2+\vert a_{31}\vert^2-\vert a_{33}-a_{31}\vert^2\Big)\Big(\vert b_{13}\vert^2+\vert b_{11}\vert^2-\vert b_{13}-b_{11}\vert^2\Big)\Bigg].
    \end{eqnarray}   
  \subsection{Phase error rate}
  We begin by calculating one component of the phase error rate in the $l^{\text{th}}$ time slot, $\langle\psi_{\text{in}}\vert E^{\dagger}\ket{a_1b_2}\bra{a_1b_2}E\otimes X_{A_{1}B_{1}}\otimes I_{A_{2}B_{2}}\vert\psi_{\text{in}}\rangle$ in this section. Similar to Eq.\eqref{eq:ber4}, we can factorize the phase error rate as,
  
   \begin{eqnarray}
   (a_{11}^*-(-1)^{j_1}&&a_{12}^*+(-1)^{j_2}a_{13}^*)(b_{21}^*-(-1)^{\tilde{j_1}}b_{22}^*+(-1)^{\tilde{j_2}}b_{23}^*)\nonumber \\
   &&
   \times\big(a_{11}+(-1)^{j_1}a_{12}+(-1)^{j_2}a_{13}\big)\big(b_{21}+(-1)^{\tilde{j_1}}b_{22}+(-1)^{\tilde{j_2}}b_{23}\big). \label{eq:phase}
   \end{eqnarray}
   We use the fact that $X$ flips the qubit $\ket{j}$, write $(-1)^{j+1}$ as $-(-1)^j$, and get the expanded value of Eq.~\eqref{eq:phase} as,
    \begin{equation}
    16\Big(\vert a_{11}\vert^2-\vert a_{12}\vert^2+\vert a_{13}\vert^2\Big)\Big(\vert b_{21}\vert^2-\vert b_{22}\vert^2+\vert b_{23}\vert^2\Big).\label{eq:phase1}
    \end{equation}
    We can also write Eq.~\eqref{eq:phase} as, 
  \begin{eqnarray}
   \big[(a_{11}^*&+&(-1)^{j_1}a_{12}^*+(-1)^{j_2}a_{13}^*)(b_{21}^*+(-1)^{\tilde{j_1}}b_{22}^*+(-1)^{\tilde{j_2}}b_{23}^*)-2((-1)^{j_1}a_{12}^*b_{21}^*\nonumber \\
   &&
   +(-1)^{j_1+\tilde{j_2}}a_{12}^*b_{23}^*+(-1)^{\tilde{j_1}}a_{11}^*b_{22}^*+(-1)^{\tilde{j_1}+j_2}a_{13}^*b_{22}^*)\big]
   \big(a_{11}+(-1)^{j_1}a_{12}\nonumber \\
   &&
   +(-1)^{j_2}a_{13}\big)\big(b_{21}+(-1)^{\tilde{j_1}}b_{22}+(-1)^{\tilde{j_2}}b_{23}\big).
   \end{eqnarray} 
   The above equation, when summed over all the possible values of $\ket{j}=\ket{j_1\tilde{j}_1j_2\tilde{j}_2}$ gives
   \begin{eqnarray}
    \langle\psi_{\text{in}}\vert E^{\dagger}&&\ket{a_1b_2}\bra{a_1b_2}E\otimes Z_{A_{1}B_{1}}\otimes I_{A_{2}B_{2}}\vert\psi_{\text{in}}\rangle-\frac{2\times 16}{2\times 144}\big[\vert a_{12}\vert^2\big(\vert b_{21}\vert^2+\vert b_{23}\vert^2\big)\nonumber \\
    &&
    +\vert b_{22}\vert^2\big(\vert a_{11}\vert^2+\vert a_{13}\vert^2\big)\big].  \label{eq:phase-2}
   \end{eqnarray}
   Calculating the remaining three terms (cf. Eqs.~\eqref{eq:per},~\eqref{eq:fil1} and ~\eqref{eq:filter2}) along the lines of Eq.~\eqref{eq:phase} and Eq.~\eqref{eq:phase1}, we get the total phase error rate  as,

\begin{eqnarray}
     1-&&\frac{16}{2\times144}\Bigg[\Big(\vert a_{11}\vert^2-\vert a_{12}\vert^2+\vert a_{13}\vert^2\Big)\Big(\vert b_{21}\vert^2-\vert b_{22}\vert^2+\vert b_{23}\vert^2\Big)+\Big(\vert a_{21}\vert^2-\vert a_{22}\vert^2+\vert a_{23}\vert^2\Big) \nonumber \\
    &&
    \times\Big(\vert b_{11}\vert^2-\vert b_{12}\vert^2+\vert b_{13}\vert^2\Big)\Big(\vert a_{11}\vert^2+\vert a_{12}\vert^2-\vert a_{13}\vert^2\Big)\Big(\vert b_{31}\vert^2+\vert b_{32}\vert^2-\vert b_{33}\vert^2\Big)\nonumber \\
    &&
    +\Big(\vert a_{31}\vert^2+\vert a_{32}\vert^2-\vert a_{33}\vert^2\Big)\Big(\vert b_{11}\vert^2+\vert b_{12}\vert^2-\vert b_{13}\vert^2\Big)\Bigg].
\end{eqnarray}
Along the lines of Eq.~\eqref{eq:phase-2}, we can express the phase error rate in terms of bit error rate as,
\begin{eqnarray}
e_{\text{p}}= e_{\text{b}}-\frac{2\times 16}{2\times 144}&&\Big[\big\{\vert a_{12}\vert^2\big(\vert b_{21}\vert^2+\vert b_{23}\vert^2\big)+\vert b_{22}\vert^2\big(\vert a_{11}\vert^2+\vert a_{13}\vert^2\big)\big\}+\big\{\vert a_{22}\vert^2\Big(\vert b_{11}\vert^2\nonumber \\
&&
+\vert b_{13}\vert^2\big)+\vert b_{12}\vert^2\big(\vert a_{21}\vert^2+\vert a_{23}\vert^2\big)\big\}+\big\{\vert a_{13}\vert^2\big(\vert b_{31}\vert^2+\vert b_{32}\vert^2\big)\nonumber\\
&&
+\vert b_{33}\vert^2\big(\vert a_{11}\vert^2+\vert a_{12}\vert^2\big)\big\}+\big\{\vert a_{33}\vert^2\big(\vert b_{11}\vert^2+\vert b_{12}\vert^2\big)\nonumber \\
&&
+\vert b_{13}\vert^2\big(\vert a_{31}\vert^2+\vert a_{32}\vert^2\big)\big\}\Big].\label{eq:phase-3}
\end{eqnarray}
From Eq.~\eqref{eq:phase-3} , we can bound the phase error in each time slot as,
\begin{equation}
 e_{\text{p}}^{(l)}\leq e_{\text{b}}^{(l)}, \forall \; l.   
\end{equation}
\section{Asymptotic key analysis of DPS-MDI}\label{sec:asymptotic}
\subsection{DPS-MDI key rate with single-photon states}
We calculate the asymptotic key rate of the single-photon source based DPS-MDI, while taking into account the effects of channel loss, background counts, and misalignment errors. We model Alice's and Bob's lossy channels as beamsplitters with transmissivity $\eta_{a}$ and $\eta_{b}$, respectively. After passing through the lossy channels, the joint input state of Alice and Bob (Eq.~\eqref{eq:dps_mdi}) appears as a mixed state to Charles, before he carries out the beamsplitter measurement :
	\begin{eqnarray} 
 	 	\rho_{\text{in}}&=&\frac{\eta_{a}\eta_{b}}{9}\ket{\psi_{11}}\bra{\psi_{11}} +\frac{\eta_{a}(1-\eta_{b})}{3}\ket{\psi_{10}}\bra{\psi_{10}}+\frac{(1-\eta_{a})\eta_{b}}{3}\ket{\psi_{01}}\bra{\psi_{01}}\nonumber \\
 	 	    &&
 	 	    +(1-\eta_{a})(1-\eta_{b})\ket{\psi_{00}}\bra{\psi_{00}},
 	 	\end{eqnarray}\label{eq:mixes_state}
 where, 
 	 \begin{align}
 	 \ket{\psi_{11}} & = \left( \, \ket{100}_{a} + e^{i\phi_{{a}_{1}}} \ket{010}_{a} + e^{i\phi_{{a}_{2}}} \ket{001}_{a} \, \right) \otimes \left( \, \ket{100}_{b} + e^{i\phi_{{b}_{1}}} \ket{010}_{b} + e^{i\phi_{{b}_{2}}} \ket{001}_{b} \, \right), \nonumber \\
 	 \ket{\psi_{10}} & = \left( \, \ket{100}_{a} + e^{i\phi_{{a}_{1}}} \ket{010}_{a} + e^{i\phi_{{a}_{2}}} \ket{001}_{a} \, \right) \otimes \ket{000}_{b},\\ 
 	 \nonumber
 	 \ket{\psi_{01}} &= \, \ket{000}_{a} \otimes \left( \, \ket{100}_{b} + e^{i\phi_{{b}_{1}}} \ket{010}_{b} + e^{i\phi_{{b}_{2}}} \ket{001}_{b} \, \right),\\ \nonumber
 	 \text{and}\qquad 	 
 	 \ket{\psi_{00}} &= \, \ket{000}_{a} \otimes \ket{000}_{b}.
 	 \end{align}
 Here, $\ket{\psi_{11}}$ corresponds to the scenario when photons from both Alice and Bob reach the measurement unit. $\ket{\psi_{10}}$ ($\ket{\psi_{01}}$) is the joint input state of Alice and Bob when Bob's (Alice's) photon gets lost in the channel, and only Alice's (Bob's) photon reaches Charles. $\ket{\psi_{00}}$ represents the case when both Alice as well as Bob's photons get lost in the channel.   
 
As per Eq.~\eqref{eq:bsm2}, the beamsplitter transforms $\ket{\psi_{11}}$ into
 \begin{eqnarray} \label{eq:bsm_output2}
	\ket{\psi_{11}}_{\text{out}}&=&\frac{1}{2}e^{i(\phi_{{b}_{1}}+\phi_{{b}_{2}})}\Big[ \, e^{i\Delta\phi_{1}}\Big(\ket{011,000}_{cd}-\ket{010,001}_{cd}+\ket{001,010}_{cd}-\ket{000,011}_{cd}\Big)\nonumber\\\nonumber \\
 	&&
    +e^{i\Delta\phi_{2}}\Big(\ket{011,000}_{cd}+\ket{010,001}_{cd} -\ket{001,010}_{cd}-\ket{000,011}_{cd}\Big)\nonumber \\
 	&&
 	+ e^{-i\phi_{{b}_{1}}}\Big\{\Big(\ket{101,000}_{cd}-\ket{100,001}_{cd}+\ket{001,100}_{cd}-\ket{000,101}_{cd}\Big)\nonumber\\
 	&&
 	+e^{i\Delta\phi_{2}}\Big(\ket{101,000}_{cd}+\ket{100,001}_{cd}-\ket{001,100}_{cd}-\ket{000,101}_{cd}\Big)\Big\}\nonumber \\
 	&&
 	+e^{-i\phi_{{b}_{2}}}\Big\{\Big(\ket{110,000}_{cd}-\ket{100,010}_{cd}+\ket{010,100}_{cd}\,-\ket{000,110}_{cd}\Big)\nonumber\\
 	&&
 	+e^{i\Delta\phi_{1}}\Big(\ket{110,000}_{cd}+\ket{100,010}_{cd} -\ket{010,100}_{cd}-\ket{000,110}_{cd}\Big)\Big\}\Big]\nonumber \\
 	&&
 	+\frac{1}{\sqrt{2}}e^{i(\phi_{{b}_{1}}+\phi_{{b}_{2}})}\Big[\Big( \,e^{-{i(\phi_{{b}_{1}}+\phi_{{b}_{2}})}}\ket{200,000}_{cd}-\ket{000,200}\Big)+e^{i(\phi_{{a}_{1}}-\phi_{{b}_{2}})} \nonumber \\
 	&&
 	\times\Big(\ket{020,000}_{cd}-\ket{000,020}_{cd}\Big)
 	+e^{i(\phi_{{a}_{2}}-\phi_{{b}_{1}})}\Big(\ket{002,000}-\ket{000,002}\Big)\Big]
 	\end{eqnarray}
Similarly, the action of beamspitter on $\ket{\psi_{10}}$, $\ket{\psi_{01}}$ and $\ket{\psi_{00}}$ are 
	\begin{eqnarray} \label{eq:1_bsm_output3}
 		\ket{\psi_{10}}_{\text{out}} & = &\frac{1}{\sqrt{2}}\Big[ \, \ket{100,000}_{cd}+\ket{000,100}_{cd}+e^{\phi_{a_{1}}}\big(\ket{010,000}_{cd}+\ket{000,010}_{cd}\big)\nonumber\\
 					& & \qquad +e^{\phi_{a_{2}}}\big(\ket{001,000}_{cd} +\ket{000,001}_{cd}\big)\Big], \\
        \label{eq:2_bsm_output3}
  		\ket{\psi_{10}}_{\text{out}} & = & \frac{1}{\sqrt{2}}\Big[ \ket{100,000}_{cd}-\ket{000,100}_{cd}+e^{\phi_{a_{1}}}\big(\ket{010,000}_{cd}-\ket{000,010}_{cd}\big)\nonumber\\
		& & \qquad +e^{\phi_{a_{2}}}\big(\ket{001,000}_{cd}-\ket{000,001}_{cd}\big)\Big], \\ 
  	 \text{and} 
  	 \quad \ket{\psi_{00}}_{\text{out}} &=&  \ket{000,000}_{cd}. \label{eq:bsm_output4}
\end{eqnarray}						
Table I shows the instances corresponding to the successful measurement events. These outcomes correspond to successful BSMs as we have mapped the DPS-MDI to an equivalent entanglement-based protocol. The yield ($Y_{11}$) for our protocol is defined as the probability of a successful measurement provided both Alice and Bob send single-photon states. Using Eqs.~\eqref{eq:mixes_state}-~\eqref{eq:bsm_output4}, we determine the probability of a successful measurement for all the cases shown in Table I as,
\begin{equation}
{Y}_{11}^{c(t_1,t_2)}=(1-p_{\text{dark}})^4\Big[\frac{\eta_{a}\eta_{b}}{18}+p_{\text{dark}}\Big(\frac{\eta_{a}+\eta_{b}}{3}-\frac{5\eta_{a}\eta_{b}}{9}\Big)+p_{\text{dark}}^{2}(1-\eta_{a})(1-\eta_{b})\Big],
\end{equation}
where, $p_\text{dark}$ is the dark count probability, and ${Y}_{11}^{c(t_1,t_2)}$  represents the probability that detector $c$ clicks in time-bins $1$ and $2$, given that both Alice and Bob send single-photon states. We use this notation to express the probability of a successful BSM for the other cases tabulated in Table I. 
\begin{eqnarray}
&&{Y}_{11}^{c(t_1,t_2)}={Y}_{11}^{c(t_1,t_3)}={Y}_{11}^{d(t_1,t_2)}={Y}_{11}^{d(t_1,t_3)}={Y}_{11}^{c(t_1),d(t_2)}={Y}_{11}^{c(t_2),d(t_1)}={Y}_{11}^{c(t_1),d(t_3)}\nonumber \\
&&
={Y}_{11}^{c(t_3),d(t_1)}.
\end{eqnarray}
Hence, the yield ($Y_{11}$) is expressed as follows:
\begin{eqnarray}\label{eq:yield}
Y_{11}&&={Y}_{11}^{c(t_1,t_2)}+{Y}_{11}^{c(t_1,t_3)}+{Y}_{11}^{d(t_1,t_2)}+{Y}_{11}^{d(t_1,t_3)}+{Y}_{11}^{c(t_1),d(t_2)}+{Y}_{11}^{c(t_2),d(t_1)}\nonumber \\
&&
+{Y}_{11}^{c(t_1),d(t_3)}+{Y}_{11}^{c(t_3),d(t_1)}\nonumber \\
&&
= 8(1-p_{\text{dark}})^4\Big[\frac{\eta_{a}\eta_{b}}{18}+p_{\text{dark}}\Big(\frac{\eta_{a}+\eta_{b}}{3}-\frac{5\eta_{a}\eta_{b}}{9}\Big)+p_{\text{dark}}^{2}(1-\eta_{a})(1-\eta_{b})\Big].
\end{eqnarray}
There are different scenarios that lead to errors in the DPS-MDI protocol. For example, an error occurs when the detector $c$ clicks in time-bins $1$ and $2$, but $\Delta\phi_{1}=\pi$. In general, an error arises when clicks corresponding to a successful partial BSM occur due to background noise, but $\Delta\phi_{i}$ (i=1, 2) is flipped (see Table I). Dark counts of the single-photon detectors primarily contribute to this background noise.  Thus, the error rate due to background noise is given by
\begin{equation}
e_{\text{b}}^{'}Y_{11}=8(1-p_{\text{dark}})^4\Big[p_{\text{dark}}\Big(\frac{\eta_{a}+\eta_{b}}{3}-\frac{5\eta_{a}\eta_{b}}{9}\Big)+p_{\text{dark}}^{2}(1-\eta_{a})(1-\eta_{b})\Big].
\end{equation}
We assume that the phase misalignment error is same for both $\Delta\phi_{1}$ and $\Delta\phi_{2}$, and denote this deviation of $\Delta\phi_{1}$ and $\Delta\phi_{2}$ by $\Delta_{\phi}$. Phase misalignment error arises due to the non-ideal nature of optical phase-locked loop and phase modulators used in the setup. Hence, considering phase misalignment errors, the total error rate is given by,
\begin{equation}    	
e_{\text{b}}Y_{11}=8(1-p_{\text{dark}})^4\Big[\frac{e_{d}\eta_{a}\eta_{b}}{18}+p_{\text{dark}}\Big(\frac{\eta_{a}+\eta_{b}}{3}-\frac{5\eta_{a}\eta_{b}}{9}\Big)+p_{\text{dark}}^{2}(1-\eta_{a})(1-\eta_{b})\Big],  
\end{equation}
where $e_{d}$ is the variance of $\Delta_{\phi}$. 
\subsection{DPS-MDI with decoy states}
Here, we calculate the parameters defined in Eq.~\eqref{eq:decoy}. We assume an infinite number of decoy states to get an accurate estimate of these parameters. Phase randomization is integral to decoy-state analysis. A coherent state is seen as a mixture of Fock states upon phase randomization. This prevents Eve from getting information from multi-photon pulses coming from WCS. Hence, Alice and Bob prepare phase randomized weak coherent states with intensities $\mu_{a}$ and $\mu_{b}$, respectively, of the form,
\begin{equation}
\ket{e^{i\theta_{a}}\sqrt{\mu_{a}}}^{(a)}\otimes\ket{e^{i\theta_{b}}\sqrt{\mu_{b}}}^{(b)}.
\end{equation}
Here, $\theta_{a}$ and $\theta_{b}$ ($\in[0,2\pi]$) are the overall randomized phases. Alice and Bob pass their coherent states through their respective delay lines. The construction of the delay line is such that a photon has an equal probability of traversing through each path of the delay line. This implies that when a coherent state $\ket{\sqrt{\mu}}$ with mean photon number $\mu$ passes through a 3-path delay line, each path has a coherent state $\ket{\sqrt{\frac{\mu_{a}}{3}}}$ with mean photon number $\frac{\mu}{3}$ traversing through it. Hence, the joint state after the coherent state passing through the delay line and the phase modulator is given as,
\begin{eqnarray}
&&\Bigg(\Big\vert e^{i\theta_{a}}\sqrt{\frac{\mu_{a}}{3}}\Big\rangle_{a_1}\Big\vert e^{i(\phi_{a_{1}}+\theta_{a})}\sqrt{\frac{\mu_{a}}{3}}\Big\rangle_{a_2}\Big\vert e^{i(\phi_{a_{2}}+\theta_{a})}\sqrt{\frac{\mu_{a}}{3}}\Big\rangle_{a_3}\Bigg)\nonumber \\
&&
\otimes\Bigg(\Big\vert e^{i\theta_{b}}\sqrt{\frac{\mu_{b}}{3}}\Big\rangle_{b_1}\Big\vert e^{i(\phi_{b_{1}}+\theta_{b})}\sqrt{\frac{\mu_{b}}{3}}\Big\rangle_{b_2}\Big\vert e^{i(\phi_{b_{2}}+\theta_{b})}\sqrt{\frac{\mu_{b}}{3}}\Big\rangle_{b_3}\Bigg).
\end{eqnarray}
Here, $\ket{\mu}_{a_1}$ represents a coherent state traversing through path 1 of Alice's delay line (see Fig.~\ref{fig1}). We model the lossy channels as beamsplitters and express the joint state arriving at Chales's beamsplitter as,
\begin{eqnarray}
&&\Bigg(\Big\vert e^{i\theta_{a}}\sqrt{\frac{\eta_{a}\mu_{a}}{3}}\Big\rangle_{a_{1}} \Big\vert e^{i(\phi_{a_{1}}+\theta_{a})}\sqrt{\frac{\eta_{a}\mu_{a}}{3}}\Big\rangle_{a_{2}}\Big\vert e^{i(\phi_{a_{2}}+\theta_{a})}\sqrt{\frac{\eta_{a}\mu_{a}}{3}}\Big\rangle_{a_{3}}\Bigg)\nonumber \\
&&
\otimes\Bigg(\Big\vert e^{i\theta_{b}}\sqrt{\frac{\eta_{b}\mu_{b}}{3}}\Big\rangle_{b_{1}}\Big\vert e^{i(\phi_{b_{1}}+\theta_{b})}\sqrt{\frac{\eta_{b}\mu_{b}}{3}}\Big\rangle_{b_{2}}\Big\vert e^{i(\phi_{b_{2}}+\theta_{b})}\sqrt{\frac{\eta_{b}\mu_{b}}{3}}\Big\rangle_{b_{3}}\Bigg).\label{eq:joint_input_BS}
\end{eqnarray}
Coherent states can also be expressed as,
\begin{equation}
\ket{\sqrt{\mu}}= D(\sqrt{\mu})\ket{0},
\end{equation}
where $D(\sqrt{\mu})$ is the displacement operator, and is given as,
\begin{equation}
D(\sqrt{\mu})=e^{(\sqrt{\mu}{a^\dagger}-\sqrt{\mu}^{*}a)}.
\end{equation} 
Here, $a$ and $a^\dagger$ are annihilation and creation operators, respectively. The beamsplitter transforms  $a^\dagger$ at the input mode as per Eq.~\eqref{eq:bsm2}. The output state of beamsplitter, when the input is Eq.~\eqref{eq:joint_input_BS} is,
\begin{eqnarray}
&&\Big\vert e^{i\theta_{a}}\sqrt{\frac{\eta_{a}\mu_{a}}{6}}+e^{i\theta_{b}}\sqrt{\frac{\eta_{b}\mu_{b}}{6}}\Big\rangle_{c_{1}}\Big\vert e^{i\theta_{a}}\sqrt{\frac{\eta_{a}\mu_{a}}{6}}-e^{i\theta_{b}}\sqrt{\frac{\eta_{b}\mu_{b}}{6}}\Big\rangle_{d_{1}}\nonumber \\
&&
\otimes\Big\vert e^{i(\phi_{a_{1}}+\theta_{a})}\sqrt{\frac{\eta_{a}\mu_{a}}{6}}+e^{i(\phi_{b_{1}}+\theta_{b})}\sqrt{\frac{\eta_{b}\mu_{b}}{6}}\Big\rangle_{c_{2}}
\Big\vert e^{i(\phi_{a_{1}}+\theta_{a})}\sqrt{\frac{\eta_{a}\mu_{a}}{6}}-e^{i(\phi_{b_{1}}+\theta_{b})}\sqrt{\frac{\eta_{b}\mu_{b}}{6}}\Big\rangle_{d_{2}}\nonumber \\
&&
\otimes\Big\vert e^{i(\phi_{a_{2}}+\theta_{a})}\sqrt{\frac{\eta_{a}\mu_{a}}{6}}+e^{i(\phi_{b_{2}}+\theta_{b})}\sqrt{\frac{\eta_{b}\mu_{b}}{6}}\Big\rangle_{c_{3}}
\Big\vert e^{i(\phi_{a_{2}}+\theta_{a})}\sqrt{\frac{\eta_{a}\mu_{a}}{6}}-e^{i(\phi_{b_{2}}+\theta_{b})}\sqrt{\frac{\eta_{b}\mu_{b}}{6}}\Big\rangle_{d_{3}}.
\end{eqnarray} 	
Here, $\ket{\sqrt{\mu}}_{c_{1}}$ denotes a coherent state of mean photon number $\mu$ hitting the detector $c$ in time-bin $t_{1}$.

Hence, the probability of a detector clicking in a time-bin is given by,
\begin{eqnarray}\label{eq:prob_click}
&& p_{c_1}=1-(1-p_{\text{dark}})\text{exp}\Big(-\Big|e^{i\theta_{a}}\sqrt{\frac{\eta_{a}\mu_{a}}{6}}+e^{i\theta_{b}}\sqrt{\frac{\eta_{b}\mu_{b}}{6}}\Big|^{2}\Big), \nonumber \\
&&
p_{d_1}=1-(1-p_{\text{dark}}) \text{exp}\Big(-\Big|e^{i\theta_{a}}\sqrt{\frac{\eta_{a}\mu_{a}}{6}}-e^{i\theta_{b}}\sqrt{\frac{\eta_{b}\mu_{b}}{6}}\Big|^{2}\Big), \nonumber \\
&&
p_{c_2}=1-(1-p_{\text{dark}})\text{exp}\Big(-\Big|e^{i(\phi_{a_{1}}+\theta_{a})}\sqrt{\frac{\eta_{a}\mu_{a}}{6}}+e^{i(\phi_{b_{1}}+\theta_{b})}\sqrt{\frac{\eta_{b}\mu_{b}}{6}}\Big|^{2}\Big), \nonumber \\
&&
p_{d_2}=1-(1-p_{\text{dark}})\text{exp}\Big(-\Big|e^{i(\phi_{a_{1}}+\theta_{a})}\sqrt{\frac{\eta_{a}\mu_{a}}{6}}-e^{i(\phi_{b_{1}}+\theta_{b})}\sqrt{\frac{\eta_{b}\mu_{b}}{6}}\Big|^{2}\Big), \nonumber \\
&&
p_{c_3}=1-(1-p_{\text{dark}})\text{exp}\Big(-\Big|e^{i(\phi_{a_{2}}+\theta_{a})}\sqrt{\frac{\eta_{a}\mu_{a}}{6}}+e^{i(\phi_{b_{2}}+\theta_{b})}\sqrt{\frac{\eta_{b}\mu_{b}}{6}}\Big|^{2}\Big), \nonumber \\
&&
p_{d_3}=1-(1-p_{\text{dark}})\text{exp}\Big(-\Big|e^{i(\phi_{a_{2}}+\theta_{a})}\sqrt{\frac{\eta_{a}\mu_{a}}{6}}-e^{i(\phi_{b_{2}}+\theta_{b})}\sqrt{\frac{\eta_{b}\mu_{b}}{6}}\Big|^{2}\Big).
\end{eqnarray}
We simplify Eq.~\eqref{eq:prob_click} by defining following relations:
\begin{eqnarray}
&&
\mu'=\eta_{a}\mu_{a}+\eta_{b}\mu_{b},\nonumber \\
&&
\Delta_{\theta}=\theta_{a}-\theta_{b},\nonumber \\
&&
x=\sqrt{\eta_{a}\mu_{a}\eta_{b}\mu_{b}}/3, \nonumber \\
&&
y=(1-p_{\text{dark}})e^{-\mu^{`}/6}.\label{eq:defined}
\end{eqnarray}
Here, $\mu'$ is the average number of photons reaching the measurement unit. $\Delta_{\theta}$ denotes the phase difference between the overall random phase applied by Alice and Bob.
Using Eq.~\eqref{eq:defined}, we simplify Eq.~\eqref{eq:prob_click} as
	\begin{equation}
	\begin{split}
  &p_{c_1}=1-ye^{-x \; cos\Delta_{\theta}},\\
  &p_{c_2}= 1-ye^{-x \; cos(\Delta_{\theta}+\Delta\phi_{1})},\\
  &p_{c_3}=1-ye^{-x \; cos(\Delta_{\theta}+\Delta\phi_{2})},
\end{split} 
\quad
\begin{split}
&p_{d_1}=1-ye^{x \; cos\Delta_{\theta}},\\
&p_{d_2}=1-ye^{x \; cos(\Delta_{\theta}+\Delta\phi_{1})},\\
&p_{d_3}=1-ye^{x \; cos(\Delta_{\theta}+\Delta\phi_{2})}.
\end{split}
\end{equation}  
$Q_{\mu_{a}\mu_{b}}$ is the overall gain when Alice and Bob, respectively, use an average photon number of $\mu_{a}$ and $\mu_{b}$, and a successful measurement occurs. We can express $Q_{\mu_{a}\mu_{b}}$ for our protocol as,
 \begin{eqnarray}
&& p_{c_1}p_{c_2}\Big(1-p_{c_3}\Big)\Big(1-p_{d_1}\Big)\Big(1-p_{d_2}\Big)\Big(1-p_{d_3}\Big)\Biggl\vert_{\Delta\phi_{1}=0,\;\Delta\phi_{2}=0\;\textrm{or}\;\pi}\nonumber\\
&&
+\,p_{c_1}\Big(1-p_{c_2}\Big)p_{c_3}\Big(1-p_{d_1}\Big)\Big(1-p_{d_2}\Big)\Big(1-p_{d_3}\Big)\Biggl\vert_{\Delta\phi_{2}=0,\;\Delta\phi_{1}=0\;\textrm{or}\;\pi}\nonumber \\
&& 
 +\Big(1-p_{c_1}\Big)\Big(1-p_{c_2}\Big)\Big(1-p_{c_3}\Big)p_{d_1}p_{d_2}\Big(1-p_{d_3}\Big)\Biggl\vert_{\Delta\phi_{1}=0,\;\Delta\phi_{2}=0\;\textrm{or}\;\pi}\nonumber \\
 &&
 +\Big(1-p_{c_1}\Big)\Big(1-p_{c_2}\Big)\Big(1-p_{c_3}\Big)p_{d_1}\Big(1-p_{d_2}\Big)p_{d_3}\Biggl\vert_{\Delta\phi_{2}=0,\;\Delta\phi_{1}=0\;\textrm{or}\;\pi} \nonumber \\
&&
+\,p_{c_1}\Big(1-p_{c_2}\Big)\Big(1-p_{c_3}\Big)\Big(1-p_{d_1}\Big)p_{d_2}\Big(1-p_{d_3}\Big)\Biggl\vert_{\Delta\phi_{1}=\pi,\;\Delta\phi_{2}=0\;\textrm{or}\;\pi} \nonumber \\
&&
+\Big(1-p_{c_1}\Big)p_{c_2}\Big(1-p_{c_3}\Big)p_{d_1}\Big(1-p_{d_2}\Big)\Big(1-p_{d_3}\Big)\Biggl\vert_{\Delta\phi_{1}=\pi,\;\Delta\phi_{2}=0\;\textrm{or}\;\pi} \nonumber \\
&&
+\,p_{c_1}\Big(1-p_{c_2}\Big)\Big(1-p_{c_3}\Big)\Big(1-p_{d_1}\Big)\Big(1-p_{d_2}\Big)p_{d_3}\Biggl\vert_{\Delta\phi_{2}=\pi,\;\Delta\phi_{1}=0\;\textrm{or}\;\pi} \nonumber \\
&&
+\Big(1-p_{c_1}\Big)\Big(1-p_{c_2}\Big)p_{c_3}p_{d_1}\Big(1-p_{d_2}\Big)\Big(1-p_{d_3}\Big)\Biggl\vert_{\Delta\phi_{2}=\pi,\;\Delta\phi_{1}=0\;\textrm{or}\;\pi}.\label{eq:decoy_gain} 
  \end{eqnarray}
We substitute Eq.~\eqref{eq:defined} in Eq.~\eqref{eq:decoy_gain}, and obtain the overall gain for a given realization of $\theta_{a}$, $\theta_{b}$ as,
\begin{equation}
Q_{\mu_{a}\mu_{b}}=4y^{4}[e^{2x\;cos\Delta_{\theta}}+e^{-2x\;cos\Delta_{\theta}}-2ye^{x\;cos\Delta_{\theta}}-2ye^{-x\;cos\Delta_{\theta}}+2y^2].\label{eq:overall_gain}
\end{equation}
We should average the overall gain obtained in Eq.~\eqref{eq:overall_gain} over the random phases $\theta_{a}$ and $\theta_{b}$. Integrating over $\Delta_{\theta}$ for Eq.~\eqref{eq:overall_gain} gives the overall gain as,
\begin{equation}
Q_{\mu_{a}\mu_{b}}=8y^4[I_{0}(2x)-2yI_{0}(x)+y^{2}]\label{eq:final_overall_gain}.
\end{equation}
Here, $I_{0}(x)$ is the modified Bessel function of the first kind.
Next, we evaluate the gain of the single-photon states ($Q_{11}$) for our protocol. $Q_{11}$ is the probability of a successful BSM, given that both Alice and Bob use weak coherent states with intensities $\mu_{a}$ and $\mu_{b}$ respectively, and send single-photon pulses. We use the Poisson distribution of photon numbers in a coherent state to obtain $Q_{11}$ as,
\begin{equation}
Q_{11}=\mu_{a}\mu_{b}e^{-\mu_a-\mu_b}Y_{11},\label{eq:gain}
\end{equation}
where $Y_{11}$ is obtained through Eq.~\eqref{eq:yield}.

The error rate in the sifted key is quantified by the overall QBER ($E_{\mu_{a}\mu_{b}}$). Error occurs in our protocol when correct set of detectors click in the right time-bins (see Table I) due to dark counts even when Alice and Bob have applied wrong $\Delta_{\phi_{i}}$ (i=1, 2). For example, clicking of detectors $c$ and $d$ when $\Delta_{\phi_{1}}=\pi$ leads to error. Hence, the overall QBER can be expressed as,
 \begin{eqnarray}
&& p_{c_1}p_{c_2}\Big(1-p_{c_3}\Big)\Big(1-p_{d_1}\Big)\Big(1-p_{d_2}\Big)\Big(1-p_{d_3}\Big)\Biggl\vert_{\Delta\phi_{1}=\pi,\;\Delta\phi_{2}=0\;\textrm{or}\;\pi}\nonumber\\
&&
+\,p_{c_1}\Big(1-p_{c_2}\Big)p_{c_3}\Big(1-p_{d_1}\Big)\Big(1-p_{d_2}\Big)\Big(1-p_{d_3}\Big)\Biggl\vert_{\Delta\phi_{2}=\pi,\;\Delta\phi_{1}=0\;\textrm{or}\;\pi}\nonumber \\
&& 
 +\Big(1-p_{c_1}\Big)\Big(1-p_{c_2}\Big)\Big(1-p_{c_3}\Big)p_{d_1}p_{d_2}\Big(1-p_{d_3}\Big)\Biggl\vert_{\Delta\phi_{1}=\pi,\;\Delta\phi_{2}=0\;\textrm{or}\;\pi}\nonumber \\
 &&
 +\Big(1-p_{c_1}\Big)\Big(1-p_{c_2}\Big)\Big(1-p_{c_3}\Big)p_{d_1}\Big(1-p_{d_2}\Big)p_{d_3}\Biggl\vert_{\Delta\phi_{2}=\pi,\;\Delta\phi_{1}=0\;\textrm{or}\;\pi} \nonumber \\
&&
+\,p_{c_1}\Big(1-p_{c_2}\Big)\Big(1-p_{c_3}\Big)\Big(1-p_{d_1}\Big)p_{d_2}\Big(1-p_{d_3}\Big)\Biggl\vert_{\Delta\phi_{1}=0,\;\Delta\phi_{2}=0\;\textrm{or}\;\pi} \nonumber \\
&&
+\Big(1-p_{c_1}\Big)p_{c_2}\Big(1-p_{c_3}\Big)p_{d_1}\Big(1-p_{d_2}\Big)\Big(1-p_{d_3}\Big)\Biggl\vert_{\Delta\phi_{1}=0,\;\Delta\phi_{2}=0\;\textrm{or}\;\pi} \nonumber \\
&&
+\,p_{c_1}\Big(1-p_{c_2}\Big)\Big(1-p_{c_3}\Big)\Big(1-p_{d_1}\Big)\Big(1-p_{d_2}\Big)p_{d_3}\Biggl\vert_{\Delta\phi_{2}=0,\;\Delta\phi_{1}=0\;\textrm{or}\;\pi} \nonumber \\
&&
+\Big(1-p_{c_1}\Big)\Big(1-p_{c_2}\Big)p_{c_3}p_{d_1}\Big(1-p_{d_2}\Big)\Big(1-p_{d_3}\Big)\Biggl\vert_{\Delta\phi_{2}=0,\;\Delta\phi_{1}=0\;\textrm{or}\;\pi}.\label{eq:decoy_error} 
  \end{eqnarray}
Substituting Eq.~\eqref{eq:defined} in Eq.~\eqref{eq:decoy_error},
\begin{equation}\label{eq:overal_qber}
E^{'}_{\mu_{a}\mu_{b}}Q_{\mu_{a}\mu_{b}}=8y^4[1-ye^{x\;cos\Delta_{\theta}}-ye^{-x\;cos\Delta_{\theta}}+y^2].
\end{equation} 
Averaging over $\Delta_{\theta}$ in Eq.~\ref{eq:overal_qber}, we get
\begin{equation}\label{eq:1_final_qber}
E^{'}_{\mu_{a}\mu_{b}}Q_{\mu_{a}\mu_{b}}=8y^4[1-2yI_{0}(x)+y^2].
\end{equation}
\subsection{Phase randomization with post selection}
As evident from Fig.~\ref{fig:qber}, phase randomization leads to a high intrinsic QBER. To reduce the QBER, Alice and Bob divide the overall phase into different splices as per Eq.~\eqref{eq:slice}. They announce the segment that they used for phase randomization while sifting. This improved data processing \cite{ma2011improved} reduces the cost of error correction (cf. Eq.~\eqref{eq:cost}).
\begin{equation}\label{eq:2_final_qber}
I_{\text{ec}}=\sum\limits_{m=0}^{N-1}Q^mfH(E^{m})
\end{equation}
We assume that Alice picks up one slice out of $N$ slices randomly and Bob always select the first phase slice. Hence, for estimation the overall gain Eq.~\eqref{eq:overall_gain} needs to be averaged over $\Delta_{\theta}$ from $\frac{m\pi}{N}$
to $\frac{(m+1)\pi}{N}$, i.e.,
\begin{equation}\label{eq:integral_gain}
Q^m=\frac{N}{\pi}\int\limits_{0}^{\pi/N}d\theta_{b}\frac{1}{\pi}\int\limits_{m\pi/N}^{(m+1)\pi/N}d\theta_{a}\times4y^{4}[e^{2x\;cos\Delta_{\theta}}+e^{-2x\;cos\Delta_{\theta}}-2ye^{x\;cos\Delta_{\theta}}-2ye^{-x\;cos\Delta_{\theta}}+2y^2].
\end{equation}
Similarly, we can write the QBER as
\begin{equation}\label{eq:integral_qber}
E^{'m}_{\mu_{a}\mu_{b}}Q^{m}_{\mu_{a}\mu_{b}}=\frac{N}{\pi}\int\limits_{0}^{\pi/N}d\theta_{b}\frac{1}{\pi}\int\limits_{m\pi/N}^{(m+1)\pi/N}d\theta_{a}\times 8y^4[1-ye^{x\;cos\Delta_{\theta}}-ye^{-x\;cos\Delta_{\theta}}+y^2].
\end{equation}
Finally, we perform a numerical integration to evaluate the QBER.
	\end{widetext}
\end{document}